\documentclass[aps,prl,longbibliography,epsfigure,twocolumn,10pt,superscriptaddress]{revtex4-2}
\usepackage[dvipsnames]{xcolor} 
\usepackage[normalem]{ulem}
\usepackage[utf8]{inputenc}
\usepackage[english]{babel}
\usepackage{amsmath, amsfonts, amssymb}
\usepackage{mathtools}
\usepackage[caption = false]{subfig}
\usepackage{graphicx,epstopdf}
\usepackage{blindtext}
\usepackage{bbm}
\usepackage{enumerate}
\usepackage{multirow}
\usepackage{physics}
\usepackage{times,txfonts}
\usepackage{siunitx}
\usepackage{tabularx}
\usepackage{tikz}
\definecolor{LinkColor}{RGB}{60,60,192} 
\usepackage[colorlinks=true,linkcolor=blue,urlcolor=LinkColor,citecolor=blue,pdfusetitle]{hyperref}
\usepackage{nicefrac} 
\usepackage{orcidlink}

\newcommand\clearrow{\global\let\rowmac\relax}
\clearrow

\allowdisplaybreaks

\usetikzlibrary{shapes.geometric, arrows}
\tikzstyle{process} = [rectangle, minimum width=3cm, minimum height=1cm, text centered, draw=white,fill=white]
\tikzstyle{arrow} = [thick,->,>=stealth]

\makeatletter
\renewcommand\@make@capt@title[2]{%
	\@ifx@empty\float@link{\@firstofone}{\expandafter\href\expandafter{\float@link}}%
	{\textrm{#1}}\@caption@fignum@sep#2\quad}%
\makeatother

\makeatletter 
\renewcommand{\fnum@figure}{\textrm{FIG.~\thefigure}}
\makeatother

\begin{document}

\title{Quantum Metrology with Quantum Limit Cycles}

\author{Zhihao Liu}
\affiliation{Institute of Quantum Precision Measurement, State Key Laboratory of Radio Frequency Heterogeneous Integration, College of Physics and Optoelectronic Engineering, Shenzhen University, Shenzhen 518060, China}

\author{Xiangyu Niu}
\affiliation{Institute of Quantum Precision Measurement, State Key Laboratory of Radio Frequency Heterogeneous Integration, College of Physics and Optoelectronic Engineering, Shenzhen University, Shenzhen 518060, China}

\author{Keren Li}
\affiliation{Institute of Quantum Precision Measurement, State Key Laboratory of Radio Frequency Heterogeneous Integration, College of Physics and Optoelectronic Engineering, Shenzhen University, Shenzhen 518060, China}
\affiliation{Quantum Science Center of Guangdong-Hong Kong-Macao Greater Bay Area (Guangdong), Shenzhen 518045, China}

\author{Ran Liu}
\email{liuran@quantumsc.cn}
\affiliation{Quantum Science Center of Guangdong-Hong Kong-Macao Greater Bay Area (Guangdong), Shenzhen 518045, China}

\author{Xiaodong Yang}
\email{yangxd@szu.edu.cn}
\affiliation{Institute of Quantum Precision Measurement, State Key Laboratory of Radio Frequency Heterogeneous Integration, College of Physics and Optoelectronic Engineering, Shenzhen University, Shenzhen 518060, China}
\affiliation{Quantum Science Center of Guangdong-Hong Kong-Macao Greater Bay Area (Guangdong), Shenzhen 518045, China}
	
\author{Jun Li}
\email{lijunquantum@szu.edu.cn}
\affiliation{Institute of Quantum Precision Measurement, State Key Laboratory of Radio Frequency Heterogeneous Integration, College of Physics and Optoelectronic Engineering, Shenzhen University, Shenzhen 518060, China}
\affiliation{Quantum Science Center of Guangdong-Hong Kong-Macao Greater Bay Area (Guangdong), Shenzhen 518045, China}
	
\date{\today}	
\begin{abstract}
 Beyond nonclassical states, emergent quantum phenomena offer new metrological resources. Quantum limit cycles, in particular, sustain long-lived oscillations that prolong coherent signal accumulation. However, quantum fluctuations may limit this advantage, leaving their metrological performance unclear.
Here, we propose a quantum limit-cycle metrology protocol for collective spins by introducing nonlinear damping into the conventional collective-decay protocol. We develop a semiclassical phase-diffusion theory to characterize its metrological performance, in good agreement with numerical simulations. The limit-cycle protocol improves estimation precision and approaches Heisenberg scaling with system size. 
We further combine limit-cycle dynamics with spin squeezing, finding that squeezing improves short-time precision, but phase diffusion ultimately dominates, leaving the asymptotic scaling unchanged.
Our results establish quantum limit cycles as a metrological resource and clarify the interplay between limit-cycle dynamics and squeezing in quantum sensing.
\end{abstract}
\maketitle

\emph{Introduction}---Quantum metrology enables parameter estimation with unprecedented precision beyond the classical limit \cite{giovannetti2004quantum,giovannettiAdvancesQuantumMetrology2011,pezzeQuantumMetrologyNonclassical2018}. Traditionally, such quantum advantages have been achieved by exploiting nonclassical properties of quantum states, including entanglement and squeezing \cite{PhysRevA.46.R6797,PhysRevLett.79.3865,Quantum.Metrology,ma2011quantum,huang2024entanglement}. 
Beyond these resources, emergent quantum phenomena and dynamical effects are explored for quantum metrology \cite{montenegro2025quantum}. 
Examples include critical enhancement near quantum phase transitions \cite{PhysRevX.8.021022,PhysRevLett.121.020402,v7mf-yh8n} and non-Hermitian exceptional points \cite{PhysRevLett.124.020501,PhysRevLett.131.160801,kmtx-7x9d}, long-lived coherence in quantum time crystals \cite{montenegroQuantumMetrologyBoundary2023,7m63-lnb8,moon2026sensing} and quantum many-body scars \cite{PRXQuantum.2.020330}, as well as memory effects in non-Markovian dynamics \cite{PhysRevLett.109.233601,PhysRevLett.127.060501,yang2024control} and decoherence suppression by quantum Zeno dynamics \cite{vs19-fwdz}. The discovery of novel quantum phenomena and dynamical processes continues to reveal new opportunities for enhanced metrology.

Recently, growing interest has emerged in quantum synchronization  \cite{PhysRevLett.111.234101,PhysRevLett.112.094102,PhysRevLett.120.163601,PhysRevLett.125.013601,tao2025noise,li2025experimental,PhysRevX.15.011010,w1bm-wjl4,PhysRevLett.121.053601} and quantum continuous time crystals \cite{kongkhambut2022observation,PhysRevLett.130.150401,wu2024dissipative,huang2025observation}, where quantum limit cycles serve as the hallmark dynamical signature. Analogous to their classical counterparts, quantum limit cycles exhibit attracting behavior towards stable orbits and give rise to a slowly decaying branch in the Liouvillian spectrum with vanishing decoherence rates \cite{PhysRevLett.134.050407}. These properties have attracted considerable theoretical interest in understanding the physical origins and mechanisms of quantum limit cycles \cite{PhysRevLett.134.050407,c5wj-4s73}, as well as interest in their potential applications in quantum technologies, such as quantum metrology. However, quantum fluctuations and noise can induce phase diffusion, potentially compromising the coherent accumulation of the sensing signal. This raises a fundamental question: Can quantum limit-cycle dynamics be harnessed to enhance quantum metrology despite the detrimental effects of phase diffusion?  Beyond this, it is intriguing to ask whether limit-cycle dynamics can be combined with conventional quantum resources, such as squeezing, to achieve complementary or synergistic metrological advantages.

\begin{figure*}[htbp]
	\centering
	\includegraphics[width=0.95\linewidth]{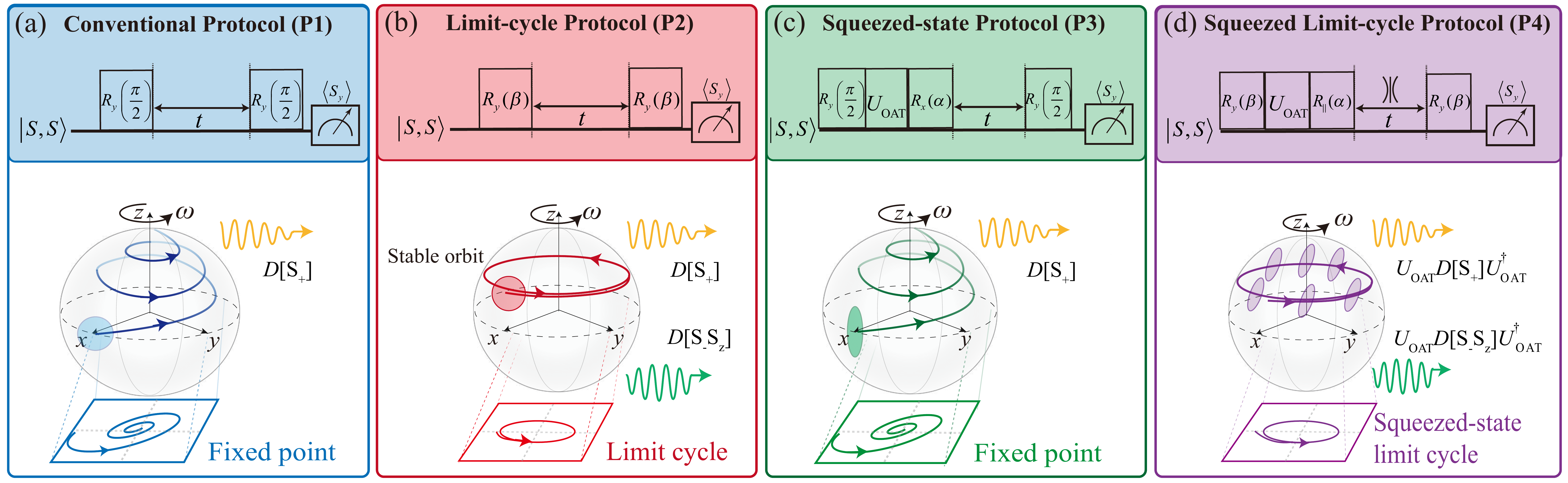}
\caption{{Schematic of different metrology protocols and their dynamics.} All protocols start from the coherent spin state $|S,S\rangle$, accumulate a phase at frequency $\omega$ during the interrogation time $t$, and read out $S_y$.
(a) Conventional protocol (P1). $R_y(\pi/2)$ prepares the probe and maps the accumulated phase onto $S_y$ for readout, while the linear dissipator $D[S_+]$ drives the probe toward a dynamical fixed point.
(b) Limit-cycle protocol (P2). $R_y(\beta)$ prepares the probe on a  limit cycle sustained by the competing dissipators $D[S_+]$ and $D[S_-S_z]$, and is reapplied for readout.
(c) Squeezed-state protocol (P3). Similar to P1, but with the probe squeezed by $U_{\rm OAT}$ and rotated by $R_x(\alpha)$ to align the minimum-variance spin component with $S_y$.
(d) Squeezed limit-cycle protocol (P4). Similar to P3, but with $R_\parallel(\alpha_{\rm opt})$ replacing $R_x(\alpha_{\rm opt})$ and the same squeezing transformation $U_{\rm OAT}$ applied to all dynamical generators, thereby driving the probe along a squeezed-state limit cycle.
}
	\label{Ramsey_LC}
\end{figure*}

Here, we address these questions by proposing a limit-cycle metrology protocol for collective spins. 
The limit-cycle dynamics is generated by introducing a nonlinear damping channel into a conventional metrology protocol with collective decay. 
We develop a phase-diffusion theory to  derive the parameter uncertainty for arbitrary system sizes, revealing two contributions: intrinsic spin-projection noise and collective phase-diffusion noise. Numerical simulations confirm that the proposed protocol sustains long-lived oscillations, enhances metrological performance, and approaches Heisenberg scaling with increasing system size, consistent with our theoretical predictions.
We further investigate combining limit-cycle dynamics with squeezing by introducing a squeezed limit-cycle protocol that incorporates squeezed states and dissipative channels. Numerical simulations show that squeezing reduces parameter uncertainty at short times when spin-projection noise dominates, but offers diminishing benefits at long times when collective phase diffusion dominates the noise. Our results provide a path to enhanced collective-spin metrology and clarify the interplay between limit-cycle dynamics and squeezing as metrological resources.

\emph{Problem formulation}---Consider estimating a parameter $\omega$ encoded in the Hamiltonian $H=\omega S_z$ of $N$ identical spin-$1/2$ particles (total spin $S=N/2$), with collective spin operators  $S_i=\frac{1}{2}\sum_{k=1}^{N}\sigma_i^{k}$ ($i \in \{x,y,z\})$.
In the conventional protocol, the probe is initialized in the coherent spin state $|S,S\rangle$, rotated by $R_y(\pi/2)$, and evolves under $H$ for an interrogation time $t$. The parameter $\omega$ is inferred from the measured expectation $\langle S_y(t)\rangle$ [Fig.~\ref{Ramsey_LC}(a)], with the metrological performance quantified by 
\begin{equation}
	\Delta\omega(t)=
	\frac{\Delta S_y(t)}
	{|\partial_{\omega}\langle S_y(t)\rangle|},
	\label{eq:error_propagation}
\end{equation}
where $\Delta S_y(t)=\sqrt{\langle S^2_y(t)\rangle -\langle S_y(t)\rangle^2}$. In realistic systems, collective spin decay limits metrological performance by driving the spin toward its ground state (a dynamical fixed point) (see Fig.~\ref{Ramsey_LC}(a)). This decay is modeled by the Lindblad operator $L_1={S_+}/{\sqrt{S}}$ with $S_+=S_x+i S_y$. The resulting dynamics is governed by $
	\dot{\rho}=-i[H,\rho]+\gamma_1D[L_1]\rho$, 
	where $\rho$ is the density operator, $D[O]\rho=O\rho O^{\dagger}-(O^{\dagger}O\rho+\rho O^{\dagger}O)/2$ is the dissipator, and $\gamma_1$ characterizes the decay rate. 
It remains challenging to counteract collective spin decay while sustaining the coherent dynamics for high-precision metrology \cite{degenQuantumSensing2017,escherGeneralFramework2011,demkowiczDobrzanskiElusiveHeisenberg2012}.

\emph{Limit-cycle protocol---}To address this challenge, we introduce a quantum limit-cycle metrology protocol, as illustrated in Fig.~\ref{Ramsey_LC}(b). The key ingredient is a nonlinear damping channel, $L_2=S_-S_z/\sqrt{S^3}$, which, together with the collective decay $L_1$, gives rise to the dynamics described by 
\begin{equation}\label{Master equation}
	\dot{\rho}=-i[H,\rho]+\gamma_1D[L_1]\rho+\gamma_2D[L_2]\rho,
\end{equation}
where $\gamma_2$ is  the nonlinear damping strength and $S_-=S_x-iS_y$.
The two dissipative channels compete: $L_1$ drives the spin toward the north-pole fixed point, while the $S_z-$dependent  channel $L_2$ provides weaker damping near the north pole and stronger damping away from it, preventing the spin from settling.
For $\gamma_2>\gamma_1$ and $S\gg 1$, this competition stabilizes a limit cycle \cite{PhysRevLett.134.050407}; see Fig.~\ref{Ramsey_LC}(b).  The resulting dynamics attracts the spin radially to the cycle while allowing free phase evolution, thereby sustaining long-lived oscillations that we exploit as a resource for quantum metrology.

Our limit-cycle protocol proceeds as illustrated in Fig.~\ref{Ramsey_LC}(b).  First, the initial state $|S,S\rangle$ is rotated onto or near the stable limit cycle via $R_y(\beta)=\exp(-i\beta S_y)$, where $\beta=\arccos\sqrt{\gamma_1/\gamma_2}$ corresponds to the polar angle of the stable cycle in the classical limit $S \rightarrow \infty$ \cite{PhysRevLett.134.050407}. 
The probe then evolves under $H=\omega S_z$ for   a time $t$ to accumulate a phase, which is subsequently rotated by $R_y(\beta)$ and read out through the expectation  $\langle S_y(t)\rangle$. 

\emph{Phase-diffusion theory of parameter uncertainty---}We now derive the parameter uncertainty for our protocol. In the classical limit $S\to\infty$, quantum fluctuations vanish and the spin follows a deterministic limit-cycle trajectory \cite{PhysRevLett.134.050407}, characterized by the normalized longitudinal coordinate $s_z^{\rm LC}=\sqrt{\gamma_1/\gamma_2}$ and transverse radius $r_{\rm LC}=\sqrt{1-(s_z^{\rm LC})^2}$. After transverse relaxation, the spin remains confined to this cycle, and its dynamics is described solely by the orbital phase. Under the signal Hamiltonian $H=\omega S_z$, the phase evolves as $\varphi(t)=\omega t+\phi_0$ (with $\phi_0$ the initial phase), providing the deterministic reference for analyzing finite-$S$ fluctuations.

In the semiclassical regime of large but finite $S$, quantum fluctuations induce both transverse and tangential fluctuations around the stable limit cycle. Transverse fluctuations rapidly restore the orbit, while tangential fluctuations accumulate, causing phase diffusion. Writing the orbital phase as $\Phi(t)=\varphi(t)+\delta\varphi(t)$, where $\varphi(t)$ is the deterministic phase and $\delta\varphi(t)$ the residual finite-$S$ fluctuation, the residual phase undergoes an effective random walk after transverse modes relax. The accumulation of small, approximately independent stochastic increments yields Gaussian diffusion with $\mathrm{Var}[\delta\varphi(t)]=2\Gamma t$, where $\Gamma$ is the finite-size phase diffusion rate. Because the phase is periodic, this Gaussian distribution is wrapped around the cycle, giving $P(\Phi,t)$ centered at $\varphi(t)$. Consequently, the quantum state approximates a statistical mixture of local wave packets distributed along the limit cycle
\begin{equation}
	\rho(t)\simeq\int_0^{2\pi}d\Phi P(\Phi,t)\rho_\Phi,
	\label{eq:phase_mixture_main}
\end{equation}
where $\rho_\Phi$ is a local quantum wave packet centered at phase $\Phi$.

\begin{figure*}[htbp]
	\centering
	\includegraphics[width=0.95\linewidth]{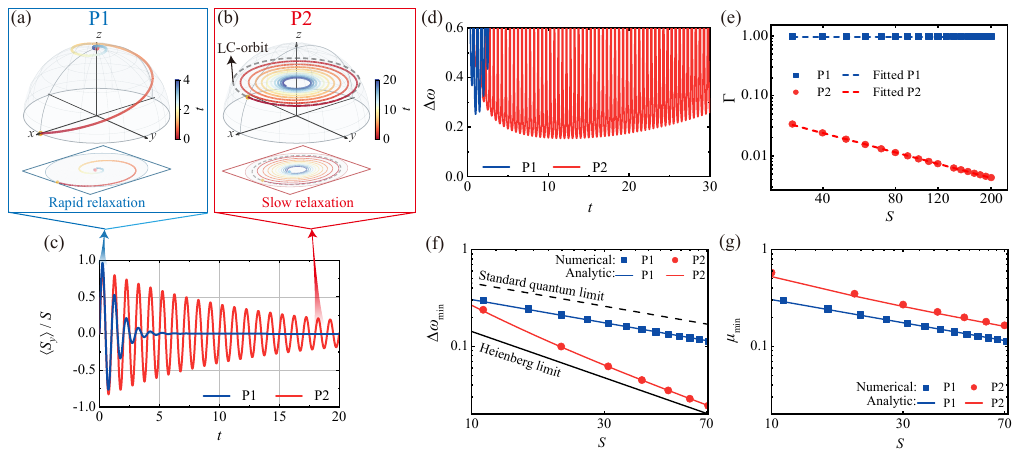}
	\caption{{Metrological performance of the limit-cycle protocol}.
		(a) Collective-spin trajectory and its $x$-$y$ projection for the conventional protocol (P1), showing rapid relaxation toward a fixed point.
		(b) Corresponding trajectory for the limit-cycle protocol (P2), showing the stable orbit and slow phase relaxation.
		(c) Normalized transverse expectation $\langle S_y(t)\rangle/S$ as a function of the encoding time $t$.
		(d) Corresponding parameter uncertainty $\Delta\omega(t)$ versus $t$.		
		(e) Phase-diffusion rate $\Gamma$ versus $S$. 
		(f) Minimum parameter uncertainty as a function of $S$. Symbols denote numerical results and solid curves are analytical predictions; the standard quantum limit and Heisenberg scaling are included as references.
		(g) Optimized sensitivity as a function of $S$, with symbols denoting numerical results and solid curves analytical predictions.
		The simulations use $\omega/\gamma_1=2\pi, \gamma_1=1$ across all panels, with $\gamma_2=4\gamma_1$ adopted for P2. The system size $S=15$ is specifically applied to (a)--(d).}
	\label{Error-Propagation}
\end{figure*}

We characterize the local wave packet $\rho_\Phi$ as a spin-coherent state centered at the corresponding limit-cycle point, with mean-spin direction $\boldsymbol{n}_\Phi=(r_{\rm LC}\cos\Phi, r_{\rm LC}\sin\Phi, s_z^{\rm LC})$. Using the first and second moments of the spin-coherent state, the local transverse signal and its intrinsic fluctuation are
\begin{equation}
\frac{\langle S_y\rangle_\Phi}{S}=r_{\rm LC}\sin\Phi, \quad
	{\rm Var}_\Phi \left(\frac{S_y}{S}\right)
	=\frac{1-r_{\rm LC}^2\sin^2\Phi}{2S}.
	\label{eq:local_phase_moments_main}
\end{equation}
The first relation follows from mean-spin direction, while the second reflects transverse fluctuations of the coherent packet.

The local spin-coherent description gives the transverse response at a fixed phase, while the observable signal is obtained by averaging this local response over the phase distribution via Eq.~\eqref{eq:phase_mixture_main}: ${\langle S_y(t)\rangle}/{S} = r_{\rm LC}\int_0^{2\pi}d\Phi P(\Phi,t)\sin\Phi= r_{\rm LC}\langle\sin\Phi(t)\rangle$, where $\langle\sin\Phi(t)\rangle$ is the first harmonic. For $\Phi(t)=\varphi(t)+\delta\varphi(t)$ with $\mathrm{Var}[\delta\varphi(t)]=2\Gamma t$, the Gaussian diffusion gives $\langle e^{ik\delta\varphi(t)}\rangle =e^{-k^2\Gamma t}$. For the first harmonic ($k=1$), we obtain $\langle\sin\Phi(t)\rangle=e^{-\Gamma t}\sin\varphi(t)$ and hence ${\langle S_y(t)\rangle}/{S}=r_{\rm LC}e^{-\Gamma t}\sin\varphi(t)$. This exponential decay arises from the broadening of the phase distribution and loss of ensemble coherence, while the radius $r_{\rm LC}$ remains unchanged.
Applying the same phase-mixture procedure to the second moment gives $\left\langle\left({S_y(t)}/{S}\right)^2\right\rangle = r_{\rm LC}^2 \langle\sin^2\Phi(t)\rangle + \left[1-r_{\rm LC}^2\langle\sin^2\Phi(t)\rangle\right]/(2S)$, with $\langle\sin^2\Phi(t)\rangle = (1-e^{-4\Gamma t}\cos[2\varphi(t)])/{2}$.

Combining the equations for the observable signal and its second moment yields
\begin{align}
{\rm Var}\left(\frac{S_y(t)}{S}\right)
& =
\frac{1}{2S}\left[1-r_{\rm LC}^2 \langle  \sin^2\Phi(t)\rangle\right] \nonumber\\
&+
r_{\rm LC}^2\left[\langle \sin^2\Phi(t)\rangle-\langle \sin\Phi(t)\rangle^2\right], 
\label{eq:Vy_main}
\end{align}
 where the first term is the intrinsic spin-projection noise and the second represents phase-diffusion noise from the displacement of local wave-packet centers. To quantify the parameter uncertainty, we combine this variance with the signal slope. At the maximum-slope points ($\varphi(t)=m\pi$), $|\partial_\omega\langle S_y/S\rangle|=r_{\rm LC}t e^{-\Gamma t}$. Substituting into Eq.~\eqref{eq:error_propagation} gives
\begin{equation}
\Delta\omega(t)
=
\frac{e^{\Gamma t}}{r_{\rm LC}t}
\sqrt{
\frac{1}{2S}
+
\frac{r_{\rm LC}^{2}}{2}
\left(1-\frac{1}{2S}\right)
\left(1-e^{-4\Gamma t}\right)
}.
\label{eq:finite_S_uncertainty_time}
\end{equation}
For $S \gg 1$ and $e^{-4\Gamma t}\ll1$, the above expression reduces to
$\Delta\omega(t)\simeq {e^{\Gamma t}}/{(\sqrt{2}t)}$.
Minimizing this expression gives the optimal interrogation time $t_{\rm opt}=1/\Gamma$ and the corresponding minimum uncertainty $
	\Delta\omega_{\min}=e\Gamma/\sqrt{2}$.
 The derivation details are provided in  Supplemental Material \cite{SM}.

\emph{Squeezed limit-cycle protocol---}To investigate whether squeezing can further enhance limit-cycle metrology, we first consider the conventional squeezed-state protocol. A coherent spin state is prepared by rotating $|S,S\rangle$ with $R_y(\pi/2)$ and squeezed via one-axis twisting $U_{\rm OAT}=\exp(-i\chi t_{\rm OAT}S_y^2)$. The minimum-variance component is aligned with $S_y$ by $R_x(\alpha_{\rm opt})$ [Fig.~\ref{Ramsey_LC}(c)], followed by parameter encoding for a time \(t\) under collective decay and readout of $\langle S_y(t)\rangle$.

%

To integrate squeezing with the quantum limit cycle, we apply the transformation to both the probe and the dynamics, yielding the protocol in Fig.~\ref{Ramsey_LC}(d). 
The probe is initialized on the cycle by $R_y(\beta)$, i.e., $\rho_\beta=R_y(\beta)|S,S\rangle\langle S,S|R_y^\dagger(\beta)$, squeezed by $U_{\rm OAT}$, and rotated by $R_\parallel(\alpha_{\rm opt})=\exp(-i\alpha_{\rm opt}S_\parallel)$ to align the minimum-variance component with $S_y$ ($S_\parallel=S_x\sin\beta+S_z\cos\beta$). The resulting state is $\sigma(0)=R_\parallel U_{\rm OAT} \rho_\beta U_{\rm OAT} ^\dagger R_\parallel^\dagger$.
To preserve the limit-cycle dynamics, we transform the generator into the frame $\sigma(t)=U_{\rm OAT}\rho(t)U_{\rm OAT}^\dagger$, giving $H'=U_{\rm OAT}HU_{\rm OAT}^\dagger$ and $L_k'=U_{\rm OAT}L_kU_{\rm OAT}^\dagger$ $(k=1,2)$. The evolution becomes $\dot{\sigma}=-i[H',\sigma]+\gamma_1\mathcal{D}[L_1']\sigma+\gamma_2\mathcal{D}[L_2']\sigma$, followed by readout of $\langle S_y(t)\rangle$.

\emph{Performance comparison between limit-cycle and conventional protocols---}We first compare both protocols at fixed $S=15$. The conventional protocol exhibits rapid decay of $\langle S_y \rangle$ toward the dynamical fixed point, whereas the limit-cycle protocol sustains oscillations over a much longer time [Fig.~\ref{Error-Propagation}(c)], as visualized by their trajectories in Figs.~\ref{Error-Propagation}(a) and (b).
The corresponding parameter uncertainties $\Delta\omega$ are shown in Fig.~\ref{Error-Propagation}(d). Setting $\gamma_1=1$ as the reference, all rates are in units of $\gamma_1$ and time in $\gamma_1^{-1}$. The conventional protocol reaches $\Delta\omega_{\min}\approx0.252$ at $t\approx1.0$, while the limit-cycle protocol achieves a much lower $\Delta\omega_{\min}\approx0.154$ at $t\approx12.0$. Thus, the prolonged oscillations extend the interrogation window and lower the optimized uncertainty at fixed system size.

We next examine the scaling of optimized uncertainty with system size $S$. From the numerical $\langle S_y\rangle$ dynamics, the phase-diffusion rate $\Gamma(S)$ for the conventional protocol remains roughly $S$-independent; see Fig.~\ref{Error-Propagation}(e). For the limit-cycle protocol, fitting over $30\leq S\leq200$ gives $\Gamma=C\gamma_1/S$ with $C=0.9453$ ($R^2=0.9924$), indicating suppressed phase diffusion and prolonged coherence at larger $S$. 
This is confirmed by the minimum uncertainty in Fig.~\ref{Error-Propagation}(f): the conventional protocol approaches the standard quantum limit, whereas the limit-cycle protocol exhibits Heisenberg scaling. Substituting $\Gamma(S)=C\gamma_1/S$ into the optimal uncertainty yields $\Delta\omega_{\min}^{\rm P2}(S)=eC\gamma_1/(\sqrt{2}S)$, consistent with our numerics. The analytical result for the conventional protocol, $\Delta\omega_{\min}^{\rm P1}(S)=\gamma_1\sqrt{\sinh 2/(4S)}$ \cite{SM}, also agrees with the data.


To account for time cost, we introduce the sensitivity \cite{degenQuantumSensing2017} $\mu(t)=\Delta\omega(t)\sqrt{t}$. From Eq.~\eqref{eq:finite_S_uncertainty_time}, the limit-cycle protocol yields $\mu(t)=\frac{e^{\Gamma t}}{r_{\rm LC}\sqrt{t}}\sqrt{\frac{1}{2S}+\frac{r_{\rm LC}^2}{2}\left(1-\frac{1}{2S}\right)\left(1-e^{-4\Gamma t}\right)}$. For $S\gg1$, the optimal time lies in the transient regime ($\Gamma t_{\rm opt}=O(S^{-1/3})\ll1$), requiring us to retain $1-e^{-4\Gamma t}$. Expanding for $\Gamma t\ll1$ yields $t_{\rm opt}\simeq\Gamma^{-1}\left[3/\left(16r_{\rm LC}^2S\left(1-\frac{1}{2S}\right)\right)\right]^{1/3}$ and $\mu_{\min}^{\rm P2}(S)\simeq\sqrt{2\Gamma}=\sqrt{2C\gamma_1/S}$ \cite{SM}. The conventional protocol similarly gives $\mu_{\min}^{\rm P1}(S)=\sqrt{\gamma_1}\sqrt{\sinh 2/(4S)}$ \cite{SM}. These results agree with our numerics; see Fig.~\ref{Error-Propagation}(g). In the large-$S$ limit, both scale as $S^{-1/2}$; thus, including the time cost removes the scaling advantage of the limit-cycle protocol.

\begin{figure}
	\centering
	\includegraphics[width=0.95\linewidth]{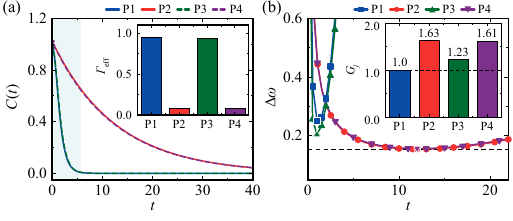}
	\caption{{Metrological performance by combining limit-cycle and squeezing.}
    (a) Normalized transverse coherence $C(t)$ for the conventional (P1), limit-cycle (P2), squeezed-state (P3), and squeezed limit-cycle (P4) protocols. The inset shows the effective decay rates $\Gamma_{\rm eff}$ extracted from exponential fits to $C(t)$. 
    (b) Parameter uncertainty $\Delta\omega(t)$ for the same protocols. The inset shows the metrological gain $G_j$ relative to P1. 
    All simulations use $S=15$ and $\omega/\gamma_1=2\pi$, with $\gamma_2/\gamma_1=0$ for P1 and P3, and $\gamma_2/\gamma_1=4$ for P2 and P4. Squeezing is applied to P3 and P4 with $\chi t_{\rm OAT}=0.05$.}
	\label{Coherent_vs_Squeezed}
\end{figure}

\emph{Interplay between limit-cycle dynamics and spin squeezing---}We now explore whether squeezing provides additional metrological benefit in the limit-cycle regime. Since its enhanced sensitivity stems from prolonged coherent oscillation, we first check if squeezing modifies the coherence lifetime. 
We characterize coherence by $C_\perp(t)=\sqrt{\langle S_x(t)\rangle^2+\langle S_y(t)\rangle^2}/S$ and remove fast carrier oscillations by averaging over one signal period, $\overline{C}_\perp(t)=T_s^{-1}\int_{t-T_s/2}^{t+T_s/2} C_\perp(t') dt'$ ($T_s=2\pi/\omega$). The normalized envelope $C(t)=\overline{C}_\perp(t)/\overline{C}_\perp(0)$ quantifies the decay. 
Figure~\ref{Coherent_vs_Squeezed}(a) shows that the limit-cycle and squeezed limit-cycle protocols exhibit nearly identical long-lived coherence, in sharp contrast to the rapid decay of the conventional and squeezed-state protocols. Exponential fits yield $\Gamma_{\rm eff}\approx0.956$ (conventional) and $\Gamma_{\rm eff}\approx0.940$ (squeezed-state). In contrast, decay rates are strongly suppressed in the limit-cycle regime, with $\Gamma_{\rm eff}\approx0.0789$ (limit-cycle) and $\Gamma_{\rm eff}\approx0.0790$ (squeezed limit-cycle); see the inset of Fig.~\ref{Coherent_vs_Squeezed}(a). Thus, squeezing has negligible effect on the coherence lifetime in both regimes.

We next quantify the metrological advantage via the gain $G_j=\Delta\omega_{\min}^{\mathrm{P1}}/\Delta\omega_{\min}^{j}$ ($j=\text{P1},\text{P2},\text{P3},\text{P4}$). As shown in Fig.~\ref{Coherent_vs_Squeezed}(b), squeezing alone reduces the minimum uncertainty from $\Delta\omega_{\min}=0.252$ to $0.205$ while keeping the optimal encoding time short ($t_{\rm opt}\approx1$). By contrast, the limit-cycle protocol achieves a lower uncertainty ($\Delta\omega_{\min}=0.154$) at a much longer time ($t_{\rm opt}\approx12$). Adding squeezing to the limit cycle yields no further improvement: the squeezed limit-cycle protocol gives $\Delta\omega_{\min}=0.156$, slightly larger than without squeezing. The corresponding gains are $G_{\rm P2}\approx1.63$, $G_{\rm P3}\approx1.23$, and $G_{\rm P4}\approx1.61$. 
These results demonstrate that squeezing and the limit cycle provide distinct, rather than additive, advantages: squeezing improves short-time uncertainty, whereas the limit cycle enhances precision by extending the useful encoding window.

\begin{figure}
	\centering
	\includegraphics[width=0.95\linewidth]{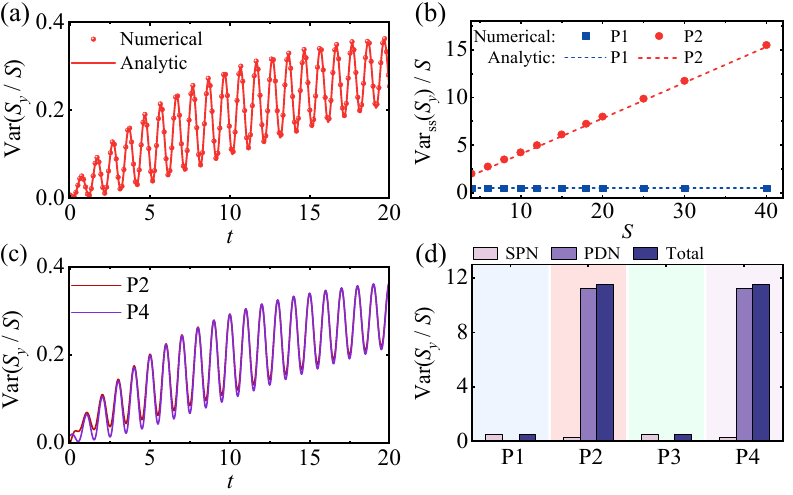}
	\caption{{Noise contributions across different metrology protocols}.
		(a) Time-dependent normalized transverse fluctuation $\mathrm{Var}(S_y/S)$ for the limit-cycle protocol. 
		(b) Steady-state fluctuation $\mathrm{Var}_{\rm ss}(S_y)/S$ versus $S$ for the conventional protocol (P1)  and the limit-cycle protocol (P2). 
		(c) Time-dependent normalized transverse fluctuation $\mathrm{Var}(S_y/S)$  for the limit-cycle protocol  and the squeezed limit-cycle protocol (P4).
		(d) Analytical decomposition of the steady-state fluctuations for all the protocols (P3: squeezed-state protocol) into spin-projection noise (SPN), phase-diffusion noise (PDN), and total noise.
		Unless otherwise specified, $S=30$, $\gamma_1=1$, and $\omega/\gamma_1=\pi$. For (a), $\Gamma=0.0334\gamma_1$; for P1 and P3, $\gamma_2/\gamma_1=0$; for P2 and P4, $\gamma_2/\gamma_1=4$; and for P4, $\chi t_{\rm OAT}=0.05$.}
	\label{fig4_fluctuations}
\end{figure}

\emph{Noise mechanisms limiting the squeezed limit-cycle protocol---}Building on our phase-diffusion theory, we use the variance decomposition in Eq.~\eqref{eq:Vy_main} to explain why squeezing provides no additional advantage in the limit-cycle regime. We validate this expression in Fig.~\ref{fig4_fluctuations}(a): the numerical variance closely matches the analytical prediction, confirming that Eq.~\eqref{eq:Vy_main} captures the relevant fluctuation dynamics.


We next examine the steady-state transverse fluctuation, $\mathrm{Var}_{\rm ss}(S_y)/S=(\langle S_y^2\rangle_{\rm ss}-\langle S_y\rangle_{\rm ss}^2)/S$ [Fig.~\ref{fig4_fluctuations}(b)]. For the conventional protocol, it remains approximately $1/2$, consistent with the projection-noise scaling $\mathrm{Var}_{\rm ss}(S_y)\propto S$. In contrast, for the limit-cycle protocol, it increases linearly with $S$, matching the steady-state prediction of Eq.~\eqref{eq:Vy_main}: $\mathrm{Var}_{\rm ss}(S_y)/S\simeq[1-r_{\rm LC}^2/2]/2+r_{\rm LC}^2S/2$. For $\gamma_2=4\gamma_1$ (where $r_{\rm LC}^2=3/4$), this reduces to $\mathrm{Var}_{\rm ss}(S_y)/S\approx5/16+3S/8$. 
This distinct scaling arises because nonlinear damping stabilizes a limit cycle with a slow phase mode: collective phase diffusion produces an $S^2$ contribution to $\mathrm{Var}_{\rm ss}(S_y)$ in the finite-size steady state, whereas the conventional protocol is limited by local spin-projection noise. Thus, the two protocols exhibit fundamentally different noise mechanisms: local projection noise dominates the conventional protocol, while collective phase-diffusion noise dominates the limit-cycle protocol.


We then examine how squeezing affects these two noise contributions and why it offers no additional advantage in the limit-cycle regime. As shown in Fig.~\ref{fig4_fluctuations}(c), squeezing reduces the short-time variance by suppressing local noise along the readout quadrature. However, the limit-cycle and squeezed limit-cycle protocols converge to the same long-time envelope, indicating that long-time fluctuations are governed by their common phase-diffusion dynamics rather than local spin-projection noise. Consequently, squeezing yields no appreciable additional gain in the limit-cycle protocol. 
The analytical decomposition via Eq.~\eqref{eq:Vy_main} makes this explicit (see Fig.~\ref{fig4_fluctuations}(d)): the conventional and squeezed-state protocols are limited by spin-projection noise, whereas phase-diffusion noise dominates both limit-cycle protocols.

\emph{Discussion---}We propose a limit-cycle metrology protocol for collective spins, demonstrating that long-lived limit-cycle oscillations enhance metrological performance. Our phase-diffusion theory shows that the optimized uncertainty approaches Heisenberg scaling, with squeezing improving short-time but not long-time performance. These results establish quantum limit-cycle dynamics and squeezing as complementary metrological resources.

Our protocol is compatible with existing platforms. The required nonlinear dissipation can be engineered by coupling the ensemble to a lossy auxiliary cavity via a magnetization-dependent Raman process; adiabatic elimination of the cavity yields the desired Lindblad channel \cite{PhysRevA.85.032111,PhysRevLett.110.120402}. Since collective cavity coupling and spin-dependent nonlinear control are well established experimentally \cite{PhysRevLett.104.073602,PhysRevA.81.021804}, our limit-cycle metrology offers a feasible pathway. More broadly, it can be extended to other continuous-variable platforms supporting self-sustained limit-cycle oscillations, such as trapped ions \cite{li2025experimental,w1bm-wjl4}, optomechanics \cite{PhysRevX.4.011015}, and superconducting circuits \cite{PhysRevLett.118.243602,PhysRevA.97.013811}.


Future work could explore tailored nonlinear dissipative channels to suppress finite-size phase diffusion. Coupled limit-cycle ensembles may induce phase synchronization \cite{PhysRevLett.113.154101,PhysRevLett.131.190402}, further enhancing long-time metrology. Phase-sensitive continuous monitoring \cite{Cabot2024ContinuousSensing,Yang2023InformationRetrieval} could also extract phase information obscured by ensemble averaging. Moreover, our nonlinear damping constitutes a form of dissipation engineering \cite{harrington2022engineered} that extends coherence times, suggesting a viable strategy to protect quantum dynamics from decoherence in quantum computing.

\emph{Acknowledgments---}\noindent
We acknowledge support by the Guangdong Provincial Quantum Science Strategic Initiative (Grant No. GDZX2505001), the National Natural Science Foundation of China (Grants No. 12675023, 12404554 and 12441502), the Innovation Program for Quantum Science and Technology (Grant No. 2024ZD0300400 and 2025ZD0300803), Guangdong Basic and Applied Basic Research Foundation (Grant No. 2026A1515010331),  the Shenzhen Science and Technology Program (Grant No. ZDYJ20251211120900001).

\emph{Data availability---}\noindent
The datasets generated during and/or analysed during the current study are available from the corresponding author on reasonable request.

\begingroup
\renewcommand{\addcontentsline}[3]{}

\endgroup

\onecolumngrid
\clearpage

\renewcommand{\baselinestretch}{1.3}\normalsize
\begin{center}
	\textbf{\large Supplemental Material for\\
		\emph{Quantum Metrology With Quantum Limit Cycles}}\\
	\vspace{0.5cm}
	Zhi-Hao Liu, Xiangyu Niu, Keren Li, Ran Liu, Xiao-Dong Yang, and Jun Li
\end{center}
\vspace{1cm}


\setcounter{table}{0}
\renewcommand{\thetable}{S\arabic{table}}%
\setcounter{figure}{0}
\renewcommand{\thefigure}{S\arabic{figure}}%
\renewcommand{\theHfigure}{S\arabic{figure}}
\setcounter{secnumdepth}{3}
\setcounter{section}{0}
\renewcommand{\thesection}{S\arabic{section}}%
\setcounter{subsection}{0}
\renewcommand{\thesubsection}{S\arabic{section}.\arabic{subsection}}
\setcounter{equation}{0}
\renewcommand{\theequation}{S\arabic{equation}}%
\setcounter{page}{1}
\renewcommand{\thepage}{SM-\arabic{page}}

\tableofcontents
\section{Mean-field dynamics and finite-\texorpdfstring{$S$}{S} decay of the transverse signal}
\label{sec:transverse-mean-spin}
This section derives the limit-cycle dynamics in the mean-field approximation and identifies the finite-$S$ transverse-coherence dynamics that motivate the phase-distribution description in the next section.
\subsection{Exact collective-spin equations}

The master equation of the limit-cycle protocol is
\begin{equation}
	\dot{\rho}
	=
	-i[\omega S_z,\rho]
	+
	\gamma_1
	D\!\left[\frac{S_+}{\sqrt S}\right]\rho
	+
	\gamma_2
	D\!\left[\frac{S_-S_z}{S^{3/2}}\right]\rho ,
	\label{eq:orig_master}
\end{equation}
where $D[L]\rho=L\rho L^\dagger-\frac12\{L^\dagger L,\rho\}$, $S$ is the total spin quantum number, and $S_\pm=S_x\pm iS_y$. The Hamiltonian $H=\omega S_z$ generates the signal-induced rotation about the $z$ axis, while the two dissipators determine the radial dynamics. For the quantum system described by the Hamiltonian $H$ and Lindblad operators $\{L_k\}$, the Heisenberg equation of motion for an arbitrary operator $O$ follows from Eq.~\eqref{eq:orig_master}, which yields
\begin{equation}
	\frac{dO}{dt}
	=
	i[\omega S_z,O]
	+
	\sum_{k=1,2}\gamma_k\,
	D^\dagger[L_k]O ,
	\label{eq:adjoint_Splus}
\end{equation}
with
\begin{equation}
	D^\dagger[L]O
	=
	L^\dagger O L
	-\frac12\{L^\dagger L,O\}
	=
	\frac12\Bigl(L^\dagger[O,L]+[L^\dagger,O]L\Bigr).
	\label{eq:Ddagger_def}
\end{equation}
For the two dissipative channels, the jump operators and their adjoints are
\begin{equation}
	L_1=\frac{S_+}{\sqrt S},\qquad
	L_2=\frac{S_-S_z}{S^{3/2}},\qquad
	L_1^\dagger=\frac{S_-}{\sqrt S},\qquad
	L_2^\dagger=\frac{S_zS_+}{S^{3/2}}.
	\label{eq:jump_ops_app}
\end{equation}
We then apply Eq.~\eqref{eq:adjoint_Splus} to the transverse operator $S_+$. 
Using the identities
$
S_-S_+
=
S(S+1)-S_z(S_z+1),\,
S_+S_-
=
S(S+1)-S_z(S_z-1),
$
and
$
S_+f(S_z)=f(S_z-1)S_+,
$
the Hamiltonian and dissipative contributions give the exact transverse equation
\begin{equation}
	\frac{dS_+}{dt}
	=
	\bigl[i\omega+A(S_z)\bigr]S_+,
	\qquad
	A(S_z)
	=
	-\frac{\gamma_1}{S}S_z
	+
	\frac{\gamma_2}{2S^3}
	\left[
	2S_z^3-S_z^2-3S_z-S(S+1)+2
	\right].
	\label{eq:Splus_A}
\end{equation}
For the longitudinal spin component $S_z$, the same adjoint equation gives
$
D^\dagger[L_1]S_z
=
\frac{1}{S}S_-S_+,
D^\dagger[L_2]S_z
=
-\frac{1}{S^3}S_z^2S_+S_-.
$
Since $[S_z,H]=0$, the Hamiltonian does not affect this component. Equation~\eqref{eq:adjoint_Splus} therefore gives
\begin{equation}
	\frac{dS_z}{dt}
	=
	\frac{\gamma_1}{S}S_-S_+
	-
	\frac{\gamma_2}{S^3}S_z^2S_+S_-.
	\label{eq:Sz-exact}
\end{equation}
Equations~\eqref{eq:Splus_A} and \eqref{eq:Sz-exact} are exact, but their expectation values involve higher-order moments, including $\langle S_zS_+\rangle$, $\langle S_z^3S_+\rangle$, and $\langle S_z^2S_+S_-\rangle$.Consequently, the equations for the first moments are not closed. To obtain a closed set of semiclassical equations, we introduce the mean-field factorization in the next subsection.

\subsection{Limit-cycle dynamics and transverse signal in the mean-field approximation}
We introduce the normalized spin components $s_\alpha=\langle S_\alpha\rangle/S$ and $s_+=s_x+is_y$. The mean-field approximation closes the moment hierarchy by factorizing normalized collective-spin moments as
$
{\langle S_{\alpha_1}\cdots S_{\alpha_n}\rangle}/{S^n}
\simeq
s_{\alpha_1}\cdots s_{\alpha_n}.
$
Applying this mean-field approximation to Eqs.~\eqref{eq:Splus_A} and \eqref{eq:Sz-exact} gives
\begin{eqnarray}
	\dot{s}_x
	&=&
	-\omega s_y
	+\left(\gamma_2s_z^3-\gamma_1s_z\right)s_x,
	\nonumber\\
	\dot{s}_y
	&=&
	\omega s_x
	+\left(\gamma_2s_z^3-\gamma_1s_z\right)s_y,
	\nonumber\\
	\dot{s}_z
	&=&
	\left(\gamma_1-\gamma_2s_z^2\right)
	\left(s_x^2+s_y^2\right).
	\label{eq:mean-field-components}
\end{eqnarray}
Equations~\eqref{eq:mean-field-components} describe the closed mean-field dynamics and conserve $s_x^2+s_y^2+s_z^2$. For the spin-coherent initial states considered here, this conserved length equals unity, so the transverse radius satisfies
$r^2=s_x^2+s_y^2=1-s_z^2.$
Using this constraint, the longitudinal equation in Eq.~\eqref{eq:mean-field-components} reduces to
$\dot{s}_z=\left(\gamma_1-\gamma_2s_z^2\right)\left(1-s_z^2\right).$ The factor $1-s_z^2$ gives the polar fixed points. A nonzero transverse radius requires $1-s_z^2>0$, in which case the remaining factor gives two nonpolar stationary branches, $s_z=\pm\sqrt{\gamma_1/\gamma_2}$.

To determine the stability of these two nonpolar branches, we linearize the longitudinal flow about each stationary solution. The positive branch is stable, whereas the negative branch is unstable. Combining the stable value of $s_z$ with the conserved spin length gives
\begin{equation}
	s_z^{\rm LC}
	=
	\sqrt{\frac{\gamma_1}{\gamma_2}},
	\qquad
	r_{\rm LC}
	=
	\sqrt{(s_x^{\rm LC})^2+(s_y^{\rm LC})^2}
	=
	\sqrt{1-\frac{\gamma_1}{\gamma_2}}.
	\label{eq:mean-field-limit-cycle}
\end{equation}
The condition $r_{\rm LC}>0$ requires $\gamma_2>\gamma_1$. On the stable branch, longitudinal perturbations decay, and conservation of the spin length converts this relaxation into restoration of the transverse radius. Equation~\eqref{eq:mean-field-limit-cycle} consequently specifies the stable orbit.

Having identified the stable orbit, we now determine the transverse motion along it, which sets the observable signal. The transverse motion is obtained by combining the first two equations in Eq.~\eqref{eq:mean-field-components}. For $s_+=s_x+i s_y$, one finds
\begin{equation}
	\dot{s}_+
	=
	\left[
	i\omega+s_z\left(\gamma_2s_z^2-\gamma_1\right)
	\right]s_+.
	\label{eq:mean-field-Splus}
\end{equation}
On the stable orbit, $s_z=s_z^{\rm LC}$ makes the real coefficient in Eq.~\eqref{eq:mean-field-Splus} vanish. Integrating the remaining phase rotation gives
\begin{equation}
	\varphi(t)=\omega t+\phi_0,
	\qquad
	s_+^{\rm MF}(t)=r_{\rm LC}e^{i\varphi(t)},
	\qquad
	s_y^{\rm MF}(t)=r_{\rm LC}\sin\varphi(t).
	\label{eq:mean-field-transverse-signal}
\end{equation}
Here, $\phi_0=\arg[s_+(0)]$ is the phase of the initial state. The mean-field signal therefore oscillates at frequency $\omega$ with a constant amplitude $r_{\rm LC}$. Mean-field theory consequently does not account for the decay of the signal contrast observed at finite $S$.

At finite $S$, taking the expectation value of Eq.~\eqref{eq:Splus_A} introduces the correlated moment $\langle A(S_z)S_+\rangle$. The mean-field factorization $\langle A(S_z)S_+\rangle\simeq A(\langle S_z\rangle)\langle S_+\rangle$ neglects correlations between $S_z$ and $S_+$ and is therefore not exact. These finite-$S$ correlations modify both the amplitude and phase of the ensemble-averaged transverse coherence. To remove the deterministic rotation generated by $H=\omega S_z$ and isolate this residual dynamics, we define
\begin{equation}
	\frac{\langle S_y(t)\rangle}{S}
	=
	\operatorname{Im}\left[e^{i\omega t}C(t)\right],
	\qquad
	C(t)= e^{-i\omega t}\frac{\langle S_+(t)\rangle}{S}.
	\label{eq:finite-S-transverse-envelope}
\end{equation}
Thus, $C(t)$ is the normalized transverse coherence in the rotating frame. Within mean-field theory, it remains constant, $C(t)=r_{\rm LC}e^{i\phi_0}$. At finite $S$, radial relaxation keeps the local wave packets close to the limit-cycle orbit, while finite-size fluctuations spread their orbital phases. Averaging over this phase distribution reduces $|C(t)|$, even though the radius of each local wave packet remains close to $r_{\rm LC}$. The decay of the ensemble-averaged signal therefore represents phase diffusion rather than radial contraction. This distinction motivates the effective phase-distribution description developed in Sec.~\ref{sec:phase-distribution}.

\section{Semiclassical phase-distribution derivation of the mean and variance of $S_y/S$}
\label{sec:phase-distribution}

Building on the finite-$S$ phase diffusion identified in Sec.~\ref{sec:transverse-mean-spin}, we develop a semiclassical phase-distribution framework to derive the mean and variance of the normalized readout $S_y/S$.
\subsection{Semiclassical phase-distribution theory}

To establish the phase-only description used below, we first contrast the restoring radial dynamics with the neutral orbital phase mode. Linearizing the mean-field flow about the stable orbit in Eq.~\eqref{eq:mean-field-limit-cycle} shows that radial deviations decay. By contrast, the dynamics is invariant under rotations about the $z$ axis. Under $U_\theta=\exp(-i\theta S_z)$, the Hamiltonian remains unchanged, whereas the jump operators transform as
$U_\theta L_1 U_\theta^\dagger=e^{-i\theta}L_1,\,U_\theta L_2 U_\theta^\dagger=e^{i\theta}L_2.$
These phase factors cancel between each jump operator and its adjoint in the corresponding Lindblad dissipator. The full master equation is therefore invariant under rotations about the $z$ axis and does not select a preferred orbital phase. In addition, previous work has established that, for the same dissipative spin model, the radial relaxation gap remains finite as $S\to\infty$, whereas the decoherence rate of the phase branch scales as $S^{-1}$ \cite{PhysRevLett.134.050407}. This spectral separation indicates that radial fluctuations relax rapidly, whereas the orbital phase remains a slow mode.

After the radial mode has relaxed, we retain only this slow phase coordinate. The deterministic orbital motion is described by the mean-field phase $\varphi(t)=\omega t+\phi_0$ in Eq.~\eqref{eq:mean-field-transverse-signal}, while $\delta\varphi(t)$ denotes the residual finite-$S$ phase fluctuation. To avoid tracking rapid microscopic fluctuations, we record the residual phase at discrete times $t_n=n\tau$. We assume that $\tau$ exceeds the correlation time of the underlying phase noise while remaining much shorter than the phase-diffusion time. 

Under this assumption, successive phase increments are treated as small, approximately independent, and stationary.
At the $n$th interval, the residual phase fluctuation is modeled as a random walk
\begin{equation}
	\delta\varphi(t_n)
	=
	\delta\varphi(t_{n-1})+\xi_n,
	\qquad
	\delta\varphi(t_n)
	=
	\sum_{j=1}^{n}\xi_j,
\end{equation}
where the second relation follows from $\delta\varphi(t_0)=0$. Rotational symmetry makes the increment statistics independent of the orbital phase. After separating the deterministic drift contained in $\varphi(t)$, we take the residual increments to have zero mean. We define the effective finite-$S$ phase-diffusion rate $\Gamma$ through their covariance:
\begin{equation}
	\langle\xi_j\rangle=0,
	\qquad
	\langle\xi_j\xi_k\rangle
	=
	2\Gamma\tau\,\delta_{jk},
	\qquad
	\langle\delta\varphi(t_n)\rangle=0,
	\qquad
	\mathrm{Var}[\delta\varphi(t_n)]
	=
	2\Gamma t_n .
	\label{eq:phase-diffusive-moments}
\end{equation}
The final relation follows from $t_n=n\tau$ and the approximate independence of the increments. It gives a phase variance that grows linearly with time and a distribution width proportional to $\sqrt{2\Gamma t}$. For $n\gg1$, the accumulated phase fluctuation approaches a Gaussian distribution, provided that the increments have finite variance and sufficiently weak correlations. A related Gaussian description of finite-size fluctuations is used for deterministic mean-field oscillator models, where the variance of the collective-order-parameter fluctuations decreases with system size \cite{Omelchenko2026FiniteSize}. Replacing the discrete time $t_n$ by the continuous variable $t$, we obtain
\begin{equation}
	P_\delta(\delta\varphi,t)
	=
	\frac{1}{\sqrt{4\pi\Gamma t}}
	\exp\left[
	-\frac{\delta\varphi^2}{4\Gamma t}
	\right],
\end{equation}
Because the orbital phase is defined modulo $2\pi$, this Gaussian distribution is understood as being wrapped around the limit-cycle orbit. The wrapping does not change averages of periodic phase harmonics. Its characteristic function is
\begin{equation}
	\left\langle e^{ik\delta\varphi(t)}\right\rangle
	=
	\exp\left(-k^2\Gamma t\right),
	\label{eq:Gaussian-phase-characteristic}
\end{equation}
so the $k$th phase harmonic decays at the rate $k^2\Gamma$. In the calculations below, the $k=1$ harmonic determines the mean transverse signal, whereas the $k=2$ harmonic determines the second moment and readout variance.
Within this phase-only description and after radial relaxation, the first-harmonic relation $\langle e^{i\delta\varphi(t)}\rangle=e^{-\Gamma t}$ gives the transverse-coherence envelope used to extract $\Gamma$. This decay represents ensemble dephasing rather than radial contraction. Individual trajectories remain close to the limit-cycle radius, while their phases spread across realizations.

Combining the deterministic mean-field phase with the stochastic phase fluctuation gives
\begin{equation}
	\Phi(t)
	=
	\varphi(t)+\delta\varphi(t).
	\label{eq:Gaussian-phase-diffusion}
\end{equation}
Since $\varphi(t)$ is deterministic, the probability distribution of the total phase $\Phi(t)=\varphi(t)+\delta\varphi(t)$ is obtained by shifting the wrapped distribution of $\delta\varphi(t)$:
$P(\Phi,t)=P_\delta^{\rm wrap}\!\left[\Phi-\varphi(t),t\right].$

\subsection{Spin-coherent-state moments at fixed phase}

To convert the phase distribution derived above into moments of the measured readout, we first derive the local moments of $S_y$ at a fixed orbital phase.
At fixed phase $\Phi$ on the limit cycle, the local mean spin direction is
\begin{equation}
	\mathbf{s}_\Phi
	=
	\left(
	r_{\rm LC}\cos\Phi,\,
	r_{\rm LC}\sin\Phi,\,
	s_z^{\rm LC}
	\right),
\end{equation}
where $s_z^{\rm LC}=\sqrt{\gamma_1/\gamma_2}$ and $r_{\rm LC}=\sqrt{1-(s_z^{\rm LC})^2}$. Since $r_{\rm LC}^2+(s_z^{\rm LC})^2=1$, the vector $\mathbf{s}_\Phi$ has unit length. For the coherent-state limit-cycle protocol analyzed here, we approximate $\rho_\Phi$ by a spin coherent state polarized along $\mathbf{s}_\Phi$. This approximation keeps the projection noise of a local coherent-state wave packet and neglects possible local squeezing corrections to the $O(S)$ noise term. Under this coherent-state approximation, the conditional first moments are
$
\langle S_i\rangle_\Phi
=
S s_{i,\Phi}.
$
In particular, for the measured transverse component, $s_{y,\Phi}=r_{\rm LC}\sin\Phi$, and hence
\begin{equation}
	\langle S_y\rangle_\Phi
	=
	S r_{\rm LC}\sin\Phi.
	\label{eq:local-Sy-mean-direct}
\end{equation}

The fixed-phase fluctuation of $S_y$ follows from the standard second-moment formula for a spin coherent state. For a spin coherent state pointing along $\mathbf{s}_\Phi$,
\begin{equation}
	\frac{1}{2}
	\langle S_iS_j+S_jS_i\rangle_\Phi
	=
	S^2s_{i,\Phi}s_{j,\Phi}
	+
	\frac{S}{2}
	\left(
	\delta_{ij}
	-
	s_{i,\Phi}s_{j,\Phi}
	\right).
	\label{eq:css-second-moment-direct}
\end{equation}
Taking $i=j=y$, Eq.~\eqref{eq:css-second-moment-direct} gives
\begin{equation}
	\langle S_y^2\rangle_\Phi
	=
	S^2s_{y,\Phi}^2
	+
	\frac{S}{2}
	\left(
	1-s_{y,\Phi}^2
	\right).
\end{equation}
Using $s_{y,\Phi}=r_{\rm LC}\sin\Phi$, we obtain
\begin{equation}
	\langle S_y^2\rangle_\Phi
	=
	S^2r_{\rm LC}^2\sin^2\Phi
	+
	\frac{S}{2}
	\left[
	1-r_{\rm LC}^2\sin^2\Phi
	\right].
	\label{eq:local-Sy2-direct}
\end{equation}

The fixed-phase variance is
\begin{equation}
	\mathrm{Var}_\Phi(S_y)
	=
	\langle S_y^2\rangle_\Phi
	-
	\langle S_y\rangle_\Phi^2.
\end{equation}
Substituting Eq.~\eqref{eq:local-Sy-mean-direct} and Eq.~\eqref{eq:local-Sy2-direct} into this definition gives

\begin{equation}
	\mathrm{Var}_\Phi(S_y)
	=
	\frac{S}{2}
	\left[
	1-r_{\rm LC}^2\sin^2\Phi
	\right].
	\label{eq:local-Sy-var}
\end{equation}
{This quantity is the local spin-projection noise at fixed phase $\Phi$. It describes the quantum fluctuation within a single local spin coherent wave packet and does not include the additional fluctuation caused by the statistical spread of the phase distribution $P(\Phi,t)$ along the limit cycle.}

\subsection{Semiclassical phase-mixture description of the density matrix}
The fixed-phase results derived above describe a local quantum wave packet at a specified point on the limit cycle. To obtain the moments measured in the full ensemble, this local quantum average must be combined with the statistical spread of the phase. In the semiclassical limit-cycle approximation, the density matrix is represented as a phase mixture of local quantum states along the orbit. The same construction underlies semiclassical phase reduction for quantum limit-cycle oscillators, where the state is reconstructed from the reduced phase distribution as a phase-weighted mixture of localized states on the classical orbit \cite{Kato2019SemiclassicalPhaseReduction}. Accordingly, we write
\begin{equation}
	\rho(t)
	\simeq
	\int d\Phi\,
	P(\Phi,t)\rho_\Phi .
	\label{eq:semiclassical-rho-decomp}
\end{equation}
Here $P(\Phi,t)$ is the effective probability distribution of the total phase at time $t$, and $\rho_\Phi$ is a local quantum wave packet centered at phase $\Phi$. All phase-averaged moments derived below are evaluated within this semiclassical phase-mixture approximation.

For any observable $O$, Eq.~\eqref{eq:semiclassical-rho-decomp} gives
\begin{equation}
	\langle O(t)\rangle
	\simeq
	\int d\Phi\,
	P(\Phi,t)
	\mathrm{Tr}\left[\rho_\Phi O\right].
	\label{eq:phase-mixture-observable}
\end{equation}
Thus the approximation keeps the two averages distinct: the quantum expectation value is first evaluated within a local state at fixed phase, and the result is then averaged over the effective phase distribution $P(\Phi,t)$. This is the rule used below for the first and second moments of $S_y$.

\subsection{First moment of \texorpdfstring{$S_y$}{Sy}}
To obtain the transverse signal, we apply Eq.~\eqref{eq:phase-mixture-observable} to the measured observable $O=S_y$. Substituting the fixed-phase mean from Eq.~\eqref{eq:local-Sy-mean-direct} then gives
\begin{equation}
	\langle S_y(t)\rangle
	\simeq
	S r_{\rm LC}M_1(t),
	\label{eq:Sy-first-moment-M1}
\end{equation}
where$M_1(t)=\int d\Phi\,P(\Phi,t)\sin\Phi$ is the first sine moment of the total phase distribution.
Using the identity
$
\sin\Phi
=
\mathrm{Im}\left(e^{i\Phi}\right),
$
we can write
\begin{equation}
	M_1(t)
	=
	\mathrm{Im}
	\left[
	\left\langle e^{i\Phi(t)}\right\rangle
	\right].
\end{equation}
The appearance of the first harmonic is fixed by the observable. At fixed
phase, the transverse coherence satisfies
$\langle S_+\rangle_\Phi\simeq S r_{\rm LC}e^{i\Phi}$, and
$\langle S_y\rangle_\Phi=\mathrm{Im}\langle S_+\rangle_\Phi$. Therefore the
mean transverse readout probes the first Fourier component
$\langle e^{i\Phi}\rangle$ of the phase distribution, rather than a higher
harmonic.

Using the effective phase-diffusion representation in Eq.~\eqref{eq:Gaussian-phase-diffusion}, the statistical average becomes
\begin{equation}
	\left\langle e^{i\Phi(t)}\right\rangle
	=
	e^{i\varphi(t)}
	\left\langle e^{i\delta\varphi(t)}\right\rangle,
\end{equation}
where $\varphi(t)$ is deterministic, so only the stochastic phase shift $\delta\varphi(t)$ is averaged over. For the first harmonic, $k=1$, Eq.~\eqref{eq:Gaussian-phase-characteristic} gives
\begin{equation}
	\left\langle e^{i\delta\varphi(t)}\right\rangle
	=
	e^{-\Gamma t},
	\qquad
	\left\langle e^{i\Phi(t)}\right\rangle
	=
	e^{-\Gamma t}e^{i\varphi(t)}.
	\label{S92}
\end{equation}
Taking the imaginary part of Eq.~\eqref{S92} gives $M_1(t)=e^{-\Gamma t}\sin\varphi(t)$, so that the first moment of the transverse readout is
\begin{equation}
	\frac{\langle S_y(t)\rangle}{S}
	\simeq
	r_{\rm LC}M_1(t)
	=
	r_{\rm LC}e^{-\Gamma t}\sin\varphi(t).
	\label{eq:normalized-Sy-mean}
\end{equation}
The factor $e^{-\Gamma t}$ is the phase-coherence factor produced by Gaussian phase diffusion. It does not describe a reduction of the radius of an individual trajectory; it arises from averaging over realizations with different stochastic phases.

\subsection{Second moment and variance of \texorpdfstring{$S_y$}{Sy}}

To determine the readout variance, we first calculate the ensemble-averaged second moment of $S_y$. We apply Eq.~\eqref{eq:phase-mixture-observable} to $O=S_y^2$ and use the fixed-phase second moment in Eq.~\eqref{eq:local-Sy2-direct}. The resulting phase dependence is described by
$
M_2(t)
\equiv
\int d\Phi\,
P(\Phi,t)\sin^2\Phi,
$
which is the sine-squared moment of the total phase distribution. The second moment is then
\begin{equation}
	\langle S_y^2(t)\rangle
	\simeq
	S^2r_{\rm LC}^2M_2(t)
	+
	\frac{S}{2}
	\left[
	1-r_{\rm LC}^2M_2(t)
	\right].
	\label{eq:Sy-second-moment-M2}
\end{equation}
It remains to evaluate $M_2(t)$ within the Gaussian phase-diffusion model. Using $\sin^2\Phi=\frac{1}{2}[1-\cos(2\Phi)]$, we obtain
$
M_2(t)
=
\frac{1}{2}
\left[
1-\langle \cos(2\Phi)\rangle
\right].
$
The second harmonic now enters for a different reason. The second moment
contains $\sin^2\Phi$, whose nonconstant phase dependence is
$\cos(2\Phi)$. Thus the phase average of $S_y^2$ probes
$\langle e^{2i\Phi}\rangle$, the $k=2$ Fourier component of the phase
distribution.

Using the phase-diffusion representation in Eq.~\eqref{eq:Gaussian-phase-diffusion}, we write
\begin{equation}
	\langle e^{2i\Phi(t)}\rangle
	=
	e^{2i\varphi(t)}
	\langle e^{2i\delta\varphi(t)}\rangle
	=e^{-4\Gamma t}e^{2i\varphi(t)}.
\end{equation}
Here Eq.~\eqref{eq:Gaussian-phase-characteristic} has been used for the second harmonic, $k=2$. Hence,
$
\langle \cos(2\Phi)\rangle
=
e^{-4\Gamma t}
\cos\left[2\varphi(t)\right].
$
Therefore, the second phase moment is
\begin{equation}
	M_2(t)
	=
	\frac{1}{2}
	\left[
	1-e^{-4\Gamma t}\cos 2\varphi(t)
	\right].
	\label{eq:M2-expression}
\end{equation}

With the phase-averaged second moment known, the total variance follows by combining it with the first moment. The variance is
$
\mathrm{Var}(S_y)
=
\langle S_y^2\rangle
-
\langle S_y\rangle^2.
$
Combining Eq.~\eqref{eq:Sy-first-moment-M1} with Eq.~\eqref{eq:Sy-second-moment-M2} gives
\begin{equation}
	{
		\mathrm{Var}(S_y)
		\simeq
		\frac{S}{2}
		\left[
		1-r_{\rm LC}^2M_2(t)
		\right]
		+
		S^2r_{\rm LC}^2
		\left[
		M_2(t)-M_1^2(t)
		\right].
	}
	\label{eq:Var-Sy-final}
\end{equation}
The two terms in Eq.~\eqref{eq:Var-Sy-final} can be traced directly to fluctuations within each local wave packet and fluctuations of the wave-packet centers. Using the phase-mixture representation in Eq.~\eqref{eq:semiclassical-rho-decomp}, Eq.~\eqref{eq:Var-Sy-final} can be rewritten as
\begin{equation}
	{
		\begin{aligned}
			\mathrm{Var}(S_y)
			&\simeq
			\int d\Phi\,P(\Phi,t)\,
			\mathrm{Var}_{\Phi}(S_y)+
			\int d\Phi\,P(\Phi,t)
			\left[
			\langle S_y\rangle_\Phi
			-
			\langle S_y(t)\rangle
			\right]^2 .
		\end{aligned}
	}
	\label{eq:variance-within-between-packets}
\end{equation}
The first integral averages the quantum variance within a local wave packet at fixed phase. Substituting Eq.~\eqref{eq:local-Sy-var} and using $M_2(t)=\langle\sin^2\Phi(t)\rangle$ gives $S[1-r_{\rm LC}^2M_2(t)]/2$, which is the first term in Eq.~\eqref{eq:Var-Sy-final}. The second integral measures the spread of the local mean values over the phase distribution. Using $\langle S_y\rangle_\Phi=Sr_{\rm LC}\sin\Phi$ and $\langle S_y(t)\rangle=Sr_{\rm LC}M_1(t)$ gives $S^2r_{\rm LC}^2[M_2(t)-M_1^2(t)]$, which is the second term in Eq.~\eqref{eq:Var-Sy-final}. Within the local spin-coherent-state approximation, we identify the first contribution as spin-projection noise and the second as phase-diffusion noise. The latter is a fluctuation of the wave-packet centers generated by the broadened orbital-phase distribution, rather than external classical noise.

For the normalized readout used in the error-propagation formula, we divide Eq.~\eqref{eq:Var-Sy-final} by $S^2$ and define
\begin{equation}
	\mathcal V_y(t)\equiv
	\mathrm{Var}\left(\frac{S_y}{S}\right)
	=
	\frac{\mathrm{Var}(S_y)}{S^2}.
	\label{eq:Var-Sy-normalized} 
\end{equation}
Substituting the expressions for $M_1(t)$ and $M_2(t)$ from Eqs.~\eqref{eq:normalized-Sy-mean} and \eqref{eq:M2-expression} into Eq.~\eqref{eq:Var-Sy-normalized} yields
\begin{equation}
	\begin{aligned}
		\mathcal V_y(t)
		&\simeq
		\frac{1}{2S}
		\left\{
		1
		-
		\frac{r_{\rm LC}^2}{2}
		\left[
		1-e^{-4\Gamma t}\cos 2\varphi(t)
		\right]
		\right\}
		+
		r_{\rm LC}^2
		\left\{
		\frac{1}{2}
		\left[
		1-e^{-4\Gamma t}\cos 2\varphi(t)
		\right]
		-
		e^{-2\Gamma t}\sin^2\varphi(t)
		\right\}.
	\end{aligned}
	\label{eq:Var-Sy-normalized-explicit}
\end{equation}
Here, $\mathcal V_{\rm SPN}(t)=\frac{1}{2S}\left[1-r_{\rm LC}^2M_2(t)\right]$ is the phase-averaged spin-projection noise (SPN) within the local wave packets, whereas $\mathcal V_{\rm PDN}(t)=r_{\rm LC}^2\left[M_2(t)-M_1^2(t)\right]$ denotes the fluctuation of the wave-packet centers generated by orbital phase diffusion noise (PDN).

\section{Minimum uncertainty and sensitivity}
\label{sec:finite-S-min-sensitivity}
According to the phase-distribution results presented in Sec.~\ref{sec:phase-distribution}, we derive the minimum parameter uncertainty $\Delta\omega_{\min}$ and the minimum  sensitivity $\mu_{\min}$, where $\mu(t)=\Delta\omega(t)\sqrt{t}$.

\subsection{Readout fluctuation and parameter uncertainty at the maximum-slope working points}
\label{sec:lc-raw-uncertainty}

The response slope of the mean signal is maximal at
$\varphi(t)=m\pi$, where $m$ is an integer. These working points form the
lower envelope of the oscillatory uncertainty curves. Setting
$\varphi(t)=m\pi$ in Eq.~\eqref{eq:Var-Sy-normalized-explicit} gives
\begin{equation}
	\mathcal V_y(t)
	=
	\frac{1}{2S}
	+
	\frac{r_{\rm LC}^2}{2}
	\left(1-\frac{1}{2S}\right)
	\left(1-e^{-4\Gamma t}\right).
	\label{eq:finite-S-working-point-variance}
\end{equation}
Equation~\eqref{eq:finite-S-working-point-variance} is the total working-point variance. At finite $S$, its second term is the net correction induced by phase spreading and should not be identified with the PDN alone. For each $S$, $\Gamma$ is obtained from the decay of the transverse signal and is held fixed during the optimization.

We first consider the uncertainty of a single interrogation without counting
the interrogation time as an additional resource. Substituting
Eq.~\eqref{eq:normalized-Sy-mean} and
Eq.~\eqref{eq:finite-S-working-point-variance} into the error-propagation
formula of the main text gives
\begin{equation}
	\Delta\omega(t)
	=
	\frac{e^{\Gamma t}}{r_{\rm LC}t}
	\sqrt{
		\frac{1}{2S}
		+
		\frac{r_{\rm LC}^2}{2}
		\left(1-\frac{1}{2S}\right)
		\left(1-e^{-4\Gamma t}\right)
	}.
	\label{eq:finite-S-raw-uncertainty}
\end{equation}

To obtain an analytical expression for the minimum uncertainty, we take $S\gg1$ and $e^{-4\Gamma t}\ll1$, which make the spin-projection contribution and the residual second-harmonic correction negligible, respectively.  The expression under the square root in Eq.~\eqref{eq:finite-S-raw-uncertainty} can therefore be approximated as
\begin{equation}
	\frac{1}{2S}
	+
	\frac{r_{\rm LC}^2}{2}
	\left(1-\frac{1}{2S}\right)
	\left(1-e^{-4\Gamma t}\right)
	\simeq
	\frac{r_{\rm LC}^2}{2}.
	\label{eq:raw-large-S-radicand}
\end{equation}
Substituting $\sqrt{r_{\rm LC}^2/2}=r_{\rm LC}/\sqrt{2}$ from Eq.~\eqref{eq:raw-large-S-radicand} into Eq.~\eqref{eq:finite-S-raw-uncertainty} cancels the factor $r_{\rm LC}$ in the denominator and gives
\begin{equation}
	\Delta\omega(t)
	\simeq
	\frac{e^{\Gamma t}}{\sqrt{2}t}.
	\label{eq:raw-large-S-uncertainty}
\end{equation}
The remaining time dependence reflects the competition between signal accumulation, represented by $1/t$, and the loss of phase coherence, represented by $e^{\Gamma t}$. Because $\Delta\omega(t)>0$, minimizing Eq.~\eqref{eq:raw-large-S-uncertainty} is equivalent to imposing the logarithmic stationary condition
\begin{equation}
	\frac{d}{dt}\ln\Delta\omega(t)
	=
	\Gamma-\frac{1}{t}
	=0,
	\qquad
	\frac{d^2}{dt^2}\ln\Delta\omega(t)
	=
	\frac{1}{t^2}>0.
	\label{eq:raw-large-S-stationary-condition}
\end{equation}
The positive second derivative establishes that the stationary point is a minimum. The optimal interrogation time is therefore
\begin{equation}
	t_{\Delta,{\rm opt}}
	=
	\frac{1}{\Gamma}.
	\label{eq:raw-large-S-optimal-time}
\end{equation}
Substituting Eq.~\eqref{eq:raw-large-S-optimal-time} into Eq.~\eqref{eq:raw-large-S-uncertainty} gives
\begin{equation}
	\Delta\omega_{\min}
	=
	\frac{e^{\Gamma t_{\Delta,{\rm opt}}}}
	{\sqrt{2}t_{\Delta,{\rm opt}}}
	=
	\frac{e\Gamma}{\sqrt{2}}.
	\label{eq:raw-large-S-minimum}
\end{equation}
At the optimal point, $e^{-4\Gamma t_{\Delta,{\rm opt}}}=e^{-4}\simeq0.018$, which validates the second-harmonic approximation. The large-$S$ reduction separately requires $1/(2S)\ll r_{\rm LC}^2/2$.

\subsection{Sensitivity in the finite-S and large-S limits}

We next include the interrogation time as a resource through
$\mu(t)=\Delta\omega(t)\sqrt{t}$.  We therefore retain the transient finite-$S$
working-point variance in Eq.~\eqref{eq:finite-S-working-point-variance}, which
gives
\begin{equation}
	\mu(t)
	=
	\frac{e^{\Gamma t}}{r_{\rm LC}\sqrt{t}}
	\sqrt{
		\frac{1}{2S}
		+
		\frac{r_{\rm LC}^2}{2}
		\left(1-\frac{1}{2S}\right)
		\left(1-e^{-4\Gamma t}\right)
	}.
	\label{eq:refined-time-sensitivity}
\end{equation}
The factor $\sqrt{t}$ accounts for the number of independent interrogations within a fixed total measurement time. It changes the temporal weighting of the uncertainty but does not modify the single-interrogation readout variance.
The transient factor $1-e^{-4\Gamma t}$ must therefore be retained during the optimization.

We introduce the dimensionless time and noise coefficients
\begin{equation}
	x=\Gamma t,
	\qquad
	A=\frac{1}{2S},
	\qquad
	B=\frac{r_{\rm LC}^2}{2}
	\left(1-\frac{1}{2S}\right).
	\label{eq:refined-time-notation}
\end{equation}
Here, $A$ is the initial projection-noise floor, whereas $B$ is the coefficient of the net variance increase induced by phase spreading. In the large-$S$ limit, $B\to r_{\rm LC}^2/2$, which is the saturation amplitude of the phase-diffusion contribution. Equation~\eqref{eq:refined-time-sensitivity} can then be written as
\begin{equation}
	\mu^2(x)
	=
	\frac{\Gamma}{r_{\rm LC}^2}F_S(x),
	\qquad
	F_S(x)
	=
	\frac{e^{2x}}{x}
	\left[A+B\left(1-e^{-4x}\right)\right].
	\label{eq:refined-time-objective}
\end{equation}
Since the prefactor $\Gamma/r_{\rm LC}^2$ is independent of $x$, the optimal interrogation time follows from the minimum of $F_S(x)$. Its logarithmic derivative gives
\begin{equation}
	0
	=
	2-\frac{1}{x_{\rm opt}}
	+
	\frac{4Be^{-4x_{\rm opt}}}
	{A+B\left(1-e^{-4x_{\rm opt}}\right)},
	\qquad
	\frac{1}{x_{\rm opt}}
	=
	2+
	\frac{4Be^{-4x_{\rm opt}}}
	{A+B\left(1-e^{-4x_{\rm opt}}\right)}.
	\label{eq:refined-time-stationary-point}
\end{equation}
For finite $S$, $F_S(x)$ diverges at both boundaries: the term $A/x$ diverges as $x\to0$, whereas the factor $e^{2x}$ produces the long-time divergence. The stationary equation therefore has a positive solution.

The large-$S$ behavior follows from $A=O(S^{-1})$ and $B=O(1)$. In this limit, $x_{\rm opt}\ll1$, and a Laurent expansion of the objective function about $x=0$ gives
\begin{equation}
	F_S(x)
	=
	\frac{A}{x}
	+4B+2A+2Ax
	+\left(\frac{4A}{3}+\frac{8B}{3}\right)x^2
	+O(x^3),
	\label{eq:refined-time-small-x-expansion}
\end{equation}
and hence
\begin{equation}
	F_S'(x)
	=
	-\frac{A}{x^2}
	+2A
	+\left(\frac{8A}{3}+\frac{16B}{3}\right)x
	+O(x^2).
	\label{eq:refined-time-small-x-derivative}
\end{equation}
The leading stationary-point balance is between $-A/x_{\rm opt}^2$ and $(16B/3)x_{\rm opt}$. Indeed, the resulting scaling $x_{\rm opt}=O(S^{-1/3})$ implies $A/x_{\rm opt}^2=O(S^{-1/3})$ and $Bx_{\rm opt}=O(S^{-1/3})$, whereas the constant term $2A=O(S^{-1})$ is subleading. The leading solution is therefore
\begin{equation}
	x_{\rm opt}
	=
	\Gamma t_{\rm opt}
	\simeq
	\left(\frac{3A}{16B}\right)^{1/3}
	=
	\left[
	\frac{3}
	{16r_{\rm LC}^2S\left(1-\frac{1}{2S}\right)}
	\right]^{1/3}
	=O(S^{-1/3}).
	\label{eq:refined-time-large-S-root}
\end{equation}
Thus, the optimum moves toward small dimensionless times as $S$ increases, $\Gamma t_{\rm opt}\to0$, and occurs before complete phase mixing.

At this stationary point, $A/x_{\rm opt}$ and $Bx_{\rm opt}^2$ are both $O(S^{-2/3})$, while the $O(S^{-1})$ correction contained in $4B$ is subleading. More explicitly,
\begin{equation}
	\mu_{\min}^2
	=
	\frac{\Gamma}{r_{\rm LC}^2}
	\left[4B+O(S^{-2/3})\right]
	=
	2\Gamma\left(1-\frac{1}{2S}\right)
	+O\!\left(\Gamma S^{-2/3}\right)
	=
	2\Gamma\left[1+O(S^{-2/3})\right].
	\label{eq:refined-time-large-S-value}
\end{equation}
{The large-$S$ optimal interrogation time and sensitivity are therefore}
\begin{equation}
	t_{\rm opt}
	\simeq
	\frac{1}{\Gamma}
	\left[
	\frac{3}
	{16r_{\rm LC}^2S\left(1-\frac{1}{2S}\right)}
	\right]^{1/3},
	\qquad
	\mu_{\min}
	\simeq
	\sqrt{2\Gamma}.
	\label{eq:refined-time-large-S-optimum}
\end{equation}
Equation~\eqref{eq:refined-time-large-S-optimum} also identifies the physical time window of the optimum. If $\Gamma=O(S^{-1})$ while the radial relaxation time $\tau_{\rm radial}=O(1)$, then $t_{\rm opt}=O(S^{2/3})$ and
\begin{equation}
	\tau_{\rm radial}\ll t_{\rm opt}\ll\Gamma^{-1}.
	\label{eq:refined-time-scale-separation}
\end{equation}
The radial degree of freedom is therefore relaxed well before the optimal readout, while the orbital phase remains only partially diffused. This separation justifies the phase-reduction description without requiring a
phase-mixed state.

The noise balance at the optimum makes this distinction explicit. From $x_{\rm opt}=O(S^{-1/3})$, $1-e^{-4x_{\rm opt}}=O(S^{-1/3})$. The phase-diffusion contribution to the variance is therefore larger than the local $O(S^{-1})$ spin-projection term, but it has not reached its $O(1)$ phase-mixed plateau. The optimized sensitivity is consequently determined by transient phase diffusion rather than by the steady-state variance. In particular, if $\Gamma=O(S^{-1})$, Eq.~\eqref{eq:refined-time-large-S-optimum} gives $\mu_{\min}=O(S^{-1/2})$, consistent with the weaker system-size scaling that results when interrogation time is counted as a resource.

This result contrasts with the parameter uncertainty in \hyperref[sec:lc-raw-uncertainty]{Sec.~S3.1}, whose optimum occurs at $\Gamma t_{\Delta,{\rm opt}}\simeq1$ and can be evaluated accurately in the phase-mixed regime. For the sensitivity, imposing $e^{-4x}\to0$ before optimization instead gives the formal stationary point $x=1/2$ and $\mu_{\min}=\sqrt{e\Gamma}$. At this point, $e^{-4x}=e^{-2}\simeq0.135$ is not negligible, and the result does not approach the true large-$S$ optimum $x_{\rm opt}=O(S^{-1/3})$. The phase-mixed limit and the large-$S$ optimization therefore do not commute.

\section{Large-S analytical derivation for the conventional protocol}
\label{sec:p1-analytical-derivation}

In this section, we derive the parameter uncertainty and sensitivity of the conventional protocol (P1) from the spin dynamics generated by the linear dissipation channel.

\subsection{Model of the conventional protocol}

The system evolves in the collective symmetric spin-$S$ subspace.
The collective operators are $S_\alpha=\sum_{k=1}^{2S}\sigma_\alpha^{(k)}/2$, with $\alpha=x,y,z$, and $S_\pm=S_x\pm iS_y$.
The signal frequency $\omega$ is encoded by the Hamiltonian
$H=\omega S_z.$
Protocol P1 contains only the linear dissipation channel, which remains active during interrogation.
Its density matrix obeys
\begin{equation}
	{
		\dot{\rho}
		=
		-i[\omega S_z,\rho]
		+
		\gamma_1D\!\left[\frac{S_+}{\sqrt S}\right]\rho,
		\qquad
		D[L]\rho
		=
		L\rho L^\dagger
		-
		\frac12\{L^\dagger L,\rho\}.
	}
	\label{eq:p1-master-equation}
\end{equation}
Thus, P1 is obtained from Eq.~\eqref{eq:orig_master} by setting $\gamma_2=0$.
The initial rotation $R_y(\pi/2)$ prepares a coherent spin state on the equator, pointing along $+x$.
Its first moments and readout variance are
\begin{equation}
	{
		\frac{\langle S_x(0)\rangle}{S}=1,
		\qquad
		\langle S_y(0)\rangle=\langle S_z(0)\rangle=0,
		\qquad
		\mathrm{Var}[S_y(0)]=\frac{S}{2}.
	}
\end{equation}
We use the dimensionless time $u=\gamma_1t$ throughout this section.
The elementary closed forms below require the large-$S$ mean-field approximation and the leading fluctuation expansion about the resulting trajectory.

\subsection{Mean trajectory and signal-response slope}
The mean trajectory is obtained by first solving for the longitudinal polarization and then using it to determine the transverse signal. For an arbitrary operator $O$, Eq.~\eqref{eq:p1-master-equation} gives the adjoint equation
\begin{equation}
	{
		\frac{d}{dt}\langle O\rangle
		=
		i\langle[\omega S_z,O]\rangle
		+
		\frac{\gamma_1}{S}
		\left\langle D^\dagger[S_+]O\right\rangle,
		\qquad
		D^\dagger[L]O
		=
		\frac12\left(L^\dagger[O,L]+[L^\dagger,O]L\right).
	}
	\label{eq:p1-adjoint-equation}
\end{equation}
To obtain the longitudinal trajectory, we set $O=S_z$ in Eq.~\eqref{eq:p1-adjoint-equation}. The Hamiltonian contribution vanishes because $[S_z,S_z]=0$. Using $[S_z,S_+]=S_+$ and $[S_-,S_z]=S_-$ in the adjoint dissipator gives
$
D^\dagger[S_+]S_z=S_-S_+.
$
Substitution into Eq.~\eqref{eq:p1-adjoint-equation} therefore yields the exact first-moment equation
\begin{equation}
	{
		\frac{d}{dt}\langle S_z\rangle
		=
		\frac{\gamma_1}{S}\langle S_-S_+\rangle .
	}
	\label{eq:p1-Sz-exact}
\end{equation}
Inserting $S_-S_+=S(S+1)-S_z^2-S_z$ into Eq.~\eqref{eq:p1-Sz-exact}, dividing by $S$, and defining $z=\langle S_z\rangle/S$ gives an equation that still contains $\langle S_z^2\rangle$.
In the large-$S$ limit, we use the mean-field approximation
$\langle S_z^2\rangle/S^2=z^2+O(S^{-1})$.
Introducing the dimensionless time $u=\gamma_1t$ then reduces the equation to
\begin{equation}
	{
		\frac{dz}{du}=1-z^2 .
	}
\end{equation}
Integrating this equation with the initial condition $z(0)=0$ gives
\begin{equation}
	{
		z(u)=\tanh u .
	}
	\label{eq:p1-z-solution}
\end{equation}

Having obtained the longitudinal trajectory, we next determine the transverse spin, whose imaginary part gives the measured signal $\langle S_y\rangle$. Setting $O=S_+$ in Eq.~\eqref{eq:p1-adjoint-equation} and using
$D^\dagger[S_+]S_+=-S_zS_+$ gives
\begin{equation}
	{
		\frac{d}{dt}\langle S_+\rangle
		=
		i\omega\langle S_+\rangle
		-
		\frac{\gamma_1}{S}\langle S_zS_+\rangle .
	}
	\label{eq:p1-Splus-exact}
\end{equation}
At leading order, $\langle S_zS_+\rangle=\langle S_z\rangle\langle S_+\rangle+O(S)$.
To separate the coherent rotation from the dissipative envelope, we write the normalized transverse spin in polar form as
\begin{equation}
	\frac{\langle S_+(t)\rangle}{S}
	=
	r_\perp(u)e^{i\omega t},
	\qquad
	r_\perp(u)
	\equiv
	\frac{|\langle S_+(t)\rangle|}{S}
	=
	\frac{\sqrt{\langle S_x(t)\rangle^2+\langle S_y(t)\rangle^2}}{S}.
\end{equation}
Here, $r_\perp$ is the normalized length of the mean spin projected onto the $x$-$y$ plane.
It determines the amplitude envelope of the signal and  quantifies the transverse coherence of P1.
The factor $e^{i\omega t}$ carries the coherent phase rotation generated by $H=\omega S_z$, whereas $r_\perp$ describes the dissipative reduction of the transverse amplitude.
Because the initial spin points along $+x$, its initial value is $r_\perp(0)=1$.
Substituting the large-$S$ factorization and the polar decomposition into Eq.~\eqref{eq:p1-Splus-exact}, the terms proportional to $i\omega$ cancel on both sides.
Dividing the remaining equation by $\gamma_1$ and using Eq.~\eqref{eq:p1-z-solution} gives
\begin{equation}
	{
		\frac{dr_\perp}{du}
		=
		-zr_\perp
		=
		-\tanh u\,r_\perp .
	}
\end{equation}
Separating variables and applying $r_\perp(0)=1$ then yields
\begin{equation}
	{
		r_\perp(u)=\operatorname{sech}u,
		\qquad
		\frac{\langle S_y(t)\rangle}{S}
		=
		\operatorname{sech}u\,\sin(\omega t).
	}
	\label{eq:p1-mean-signal}
\end{equation}
The linear dissipation channel therefore drives the spin toward the north pole while reducing the transverse amplitude nonexponentially.

The dissipative envelope is independent of a local variation of $\omega$ because the dynamics is phase covariant.
Differentiating the mean signal in Eq.~\eqref{eq:p1-mean-signal} with respect to $\omega$ at fixed $t$ therefore gives
\begin{equation}
	{
		\frac{\partial}{\partial\omega}
		\frac{\langle S_y(t)\rangle}{S}
		=
		t\operatorname{sech}u\,\cos(\omega t).
	}
	\label{eq:p1-signal-slope}
\end{equation}
For the fixed readout $S_y$, the local maximum-slope working points satisfy $\omega t=n\pi$.
At these points, the slope magnitude is $t\operatorname{sech}u$ and $\langle S_y\rangle=0$.

\subsection{Time-dependent readout fluctuation}
The mean signal alone is insufficient to determine the estimation uncertainty.
The readout fluctuation must be evaluated at the same interrogation time as the response slope.
At a maximum-slope point, the laboratory $y$ axis coincides with the local transverse direction in the frame rotating with $H$.
We therefore define
\begin{equation}
	{
		v_y(u)=\frac{\mathrm{Var}(S_y)}{S},
		\qquad
		v_y(0)=\frac12 .
	}
\end{equation}
The second moment follows from the same adjoint equation.
Setting $O=S_y^2$ in Eq.~\eqref{eq:p1-adjoint-equation} and using $[S_y,S_+]=-iS_z$ and $[S_-,S_y]=iS_z$ gives
$D^\dagger[S_+]S_y^2=S_z^2-S_y^2-\frac12\left(S_zS_y^2+2S_yS_zS_y+S_y^2S_z\right).$
At a maximum-slope point, the Hamiltonian merely defines the rotating frame and does not change the local transverse variance. Substituting the dissipator above into Eq.~\eqref{eq:p1-adjoint-equation} therefore gives
\begin{equation}
	{
		\frac{d}{dt}\langle S_y^2\rangle
		=
		\frac{\gamma_1}{S}
		\langle
		S_z^2-S_y^2
		-
		\frac12
		(
		S_zS_y^2+2S_yS_zS_y+S_y^2S_z
		)
		\rangle .
	}
	\label{eq:p1-Sy2-exact}
\end{equation}
This equation is exact, but it is not closed because it contains third-order moments.
The leading large-$S$ expansion uses
\begin{align}
	\langle S_z^2\rangle
	&=S^2z^2+O(S),\\
	\langle S_y^2\rangle
	&=Sv_y,\\
	\left\langle
	S_zS_y^2+2S_yS_zS_y+S_y^2S_z
	\right\rangle
	&=4S^2zv_y+O(S).
\end{align}
Substituting these three leading-order expressions into Eq.~\eqref{eq:p1-Sy2-exact}, using $\langle S_y\rangle=0$ at the working point, and dividing the result by $S$ gives
\begin{equation}
	{
		\frac{dv_y}{dt}
		=
		\gamma_1\left(z^2-2zv_y\right)+O(S^{-1}).
	}
\end{equation}
Keeping the leading term, replacing $z$ by Eq.~\eqref{eq:p1-z-solution}, and changing from $t$ to $u$ then gives the closed fluctuation equation
\begin{equation}
	{
		\frac{dv_y}{du}
		+2\tanh u\,v_y
		=
		\tanh^2u .
	}
	\label{eq:p1-vy-equation}
\end{equation}
Equation~\eqref{eq:p1-vy-equation} is a first-order linear differential equation with integrating factor
$\exp\!(\int 2\tanh u\,du)=\cosh^2u .$
Multiplying Eq.~\eqref{eq:p1-vy-equation} by this factor reduces it to
${d}[v_y(u)\cosh^2u]/{du}=\sinh^2u .$

Integrating both sides from $0$ to $u$, using $v_y(0)=1/2$, and evaluating
$\int_0^u\sinh^2u'\,du'=\sinh(2u)/4-u/2$ gives
\begin{equation}
	{
		v_y(u)
		=
		\frac{1}{\cosh^2u}
		\left(
		\frac12+\frac14\sinh 2u-\frac{u}{2}
		\right).
	}
\end{equation}
Finally, using $\mathrm{Var}(S_y/S)=\mathrm{Var}(S_y)/S^2=v_y/S$ converts this result into the variance of the normalized readout:
\begin{equation}
	{
		\mathcal V_y^{\rm P1}(u)
		\equiv
		\mathrm{Var}\!\left(\frac{S_y}{S}\right)
		=
		\frac{1}{S\cosh^2u}
		\left(
		\frac12+\frac14\sinh 2u-\frac{u}{2}
		\right).
	}
	\label{eq:p1-normalized-variance}
\end{equation}
Equation~\eqref{eq:p1-normalized-variance} retains the transient noise generated during the dissipative evolution.
Replacing it by the initial value $1/(2S)$ is not valid near the optimal interrogation time.

\vspace{-0.5\baselineskip}
\subsection{Optimal uncertainty and sensitivity}
\vspace{-0.15\baselineskip}
For the measured observable $S_y/S$, the parameter uncertainty follows from the error-propagation formula
\begin{equation}
	{
		\Delta\omega(t)
		=
		\frac{\sqrt{\mathcal V_y(t)}}
		{\left|\partial_\omega\langle S_y(t)/S\rangle\right|}.
	}
	\label{eq:p1-error-propagation}
\end{equation}
This expression assumes local linear response and an approximately unbiased estimator around the chosen working point.
At a maximum-slope point, substituting Eqs.~\eqref{eq:p1-signal-slope} and \eqref{eq:p1-normalized-variance} into Eq.~\eqref{eq:p1-error-propagation} first gives
\begin{equation}
	{
		\Delta\omega^{\rm P1}(u)
		=
		\frac{1}{t\operatorname{sech}u}
		\sqrt{
			\frac{1}{S\cosh^2u}
			\left(
			\frac{1}{2}+\frac{1}{4}\sinh 2u-\frac{u}{2}
			\right)
		}.
	}
\end{equation}
Using $\operatorname{sech}u=1/\cosh u$ cancels the envelope factors in the numerator and denominator.
Then substituting $t=u/\gamma_1$ and defining $A(u)=1+\sinh(2u)/2-u$ gives
\begin{equation}
	{
		\Delta\omega^{\rm P1}(u)
		=
		\frac{\gamma_1}{\sqrt{2S}}
		\frac{\sqrt{A(u)}}{u},
		\qquad
		A(u)
		=
		1+\frac12\sinh 2u-u.
	}
	\label{eq:p1-time-dependent-uncertainty}
\end{equation}
Equation~\eqref{eq:p1-time-dependent-uncertainty} is the parameter uncertainty of a single interrogation and does not count the interrogation time as an additional resource.

For a fixed total experimental duration, the number of independent repetitions scales as $1/t$.
The sensitivity used in the main text is obtained by multiplying Eq.~\eqref{eq:p1-time-dependent-uncertainty} by $\sqrt{t}=\sqrt{u/\gamma_1}$:
\begin{equation}
	{
		\mu^{\rm P1}(u)
		=
		\Delta\omega^{\rm P1}(u)\sqrt t
		=
		\sqrt{\frac{\gamma_1}{2S}}
		\sqrt{\frac{A(u)}{u}}.
	}
	\label{eq:p1-time-dependent-sensitivity}
\end{equation}
The factor $\sqrt t$ changes the time weighting but does not alter the fluctuation entering Eq.~\eqref{eq:p1-error-propagation}.

To obtain the optimized values shown in Fig.~2 of the main text, we use the fixed readout $S_y$ and $\omega/\gamma_1=2\pi$.
Substituting this frequency into the maximum-slope condition $\omega t=n\pi$ and using $u=\gamma_1t$ restricts the admissible times to
\begin{equation}
	{
		t_n=\frac{n\pi}{\omega},
		\qquad
		u_n=\gamma_1t_n=\frac{n}{2},
		\qquad n=1,2,\ldots .
	}
	\label{eq:p1-discrete-working-points}
\end{equation}
At the first three working points, $u_n=0.5,1,1.5$, the dimensionless coefficients of $\Delta\omega^{\rm P1}$ are $2.085$, $1.346$, and $1.415$.
The corresponding coefficients of $\mu^{\rm P1}$ are $1.474$, $1.346$, and $1.733$.
Both metrics therefore reach their discrete minima at $u=1$.
Since $A(1)=\sinh(2)/2$, evaluating Eqs.~\eqref{eq:p1-time-dependent-uncertainty} and \eqref{eq:p1-time-dependent-sensitivity} at $u=1$, together with $u=\gamma_1t$, gives the optimal interrogation time and the corresponding minimum values
\begin{equation}
	{
		t_{\rm opt}
		=
		\frac{1}{\gamma_1},
		\qquad
		\Delta\omega_{\min}^{\rm P1}(S)
		=
		\gamma_1\sqrt{\frac{\sinh 2}{4S}},
		\qquad
		\mu_{\min}^{\rm P1}(S)
		=
		\sqrt{\gamma_1}\sqrt{\frac{\sinh 2}{4S}}.
	}
	\label{eq:p1-minimum-uncertainty-sensitivity}
\end{equation}
Equivalently, their dimensionless coefficients satisfy ${\Delta\omega_{\min}^{\rm P1}}/{\gamma_1}={\mu_{\min}^{\rm P1}}/{\sqrt{\gamma_1}}=\sqrt{{\sinh 2}/{4S}}.$
In the units $\gamma_1=1$ used in the numerical calculation, the two optimized values are numerically identical because $t_{\rm opt}=1$.

These analytical expressions require P1 with $\gamma_2=0$, collective symmetric dynamics, the linear dissipation channel active throughout interrogation, and the large-$S$ mean-field and leading-fluctuation approximations.
At small $S$, the moment hierarchy does not close, so the finite-dimensional master equation must be solved numerically.
Error propagation further assumes local linear response and an approximately unbiased estimator near a maximum-slope working point.
Under these conditions, both quantities scale as $S^{-1/2}$, consistent with the standard-quantum-limit dependence on particle number.

\section{Steady-state fluctuation scaling and phase-diffusion contribution}
\label{sec:ss_fluctuation_criterion}
In this section, to compare the steady-state transverse fluctuations of the conventional and limit-cycle protocols in Figs.~4(b) and 4(d) of the main text, we evaluate $\mathrm{Var}_{\rm ss}(S_y)/S$ for each type of dynamics.

For the conventional protocol (P1), the linear dissipation channel drives the spin toward the polar steady state $|S,S\rangle$.
Taking the long-time limit of Eq.~\eqref{eq:p1-normalized-variance} gives
\begin{equation}
	\mathcal V_{y,\rm ss}^{\rm P1}
	=
	\lim_{u\to\infty}
	\mathcal V_y^{\rm P1}(u)
	=
	\frac{1}{2S}.
	\label{eq:p1-ss-normalized-variance}
\end{equation}
Since $\mathcal V_y=\mathrm{Var}(S_y)/S^2$, the corresponding steady-state transverse fluctuation is
\begin{equation}
	\frac{\mathrm{Var}_{\rm ss}^{\rm P1}(S_y)}{S}
	=
	S\mathcal V_{y,\rm ss}^{\rm P1}
	=
	\frac12 .
	\label{eq:standard-ss-fluctuation}
\end{equation}
This result also follows directly from the transverse variance of the polar spin-coherent state $|S,S\rangle$.
Because this steady state has neither a limit-cycle orbit nor a diffusive orbital phase, its transverse fluctuation contains only spin-projection noise:
\begin{equation}
	\mathcal V_{{\rm SPN},\rm ss}^{\rm P1}
	=
	\frac{1}{2S},
	\qquad
	\mathcal V_{{\rm PDN},\rm ss}^{\rm P1}
	=
	0 .
\end{equation}
This projection-noise value agrees with the steady-state plateau of P1 in Figs.~4(b) and 4(d) of the main text.
The squeezed-state protocol (P3) reaches the same polar steady state under the linear dissipation channel and therefore also has no steady-state phase-diffusion contribution.

For the limit-cycle protocol (P2), the steady state is phase mixed along an orbit with transverse radius $r_{\rm LC}$.
In the long-time limit, $M_1(t)\to0$ and $M_2(t)\to1/2$.
Substituting these limits into Eq.~\eqref{eq:Var-Sy-final} gives
\begin{equation}
	\mathrm{Var}_{\rm ss}(S_y)
	\simeq
	\frac{S}{2}
	\left(
	1-\frac{r_{\rm LC}^2}{2}
	\right)
	+
	\frac{S^2r_{\rm LC}^2}{2}.
	\label{eq:ss_var_Sy}
\end{equation}
Dividing by $S$ gives the quantity plotted in Fig.~4(b) of the main text:
\begin{equation}
	\frac{\mathrm{Var}_{\rm ss}(S_y)}{S}
	\simeq
	\frac12
	\left(
	1-\frac{r_{\rm LC}^2}{2}
	\right)
	+
	\frac{r_{\rm LC}^2}{2}S.
	\label{eq:ss_Fy_scaling}
\end{equation}
For the parameters used in the main text, $\gamma_2=4\gamma_1$ and hence $r_{\rm LC}^2=3/4$.
Equation~\eqref{eq:ss_Fy_scaling} reduces to
\begin{equation}
	\frac{\mathrm{Var}_{\rm ss}(S_y)}{S}
	\simeq
	\frac{5}{16}
	+
	\frac{3}{8}S.
	\label{eq:ss_Fy_main_parameters}
\end{equation}
This is the steady-state scaling used to interpret the linear increase in Fig.~4(b) of the main text.
The constant term $5/16$ is the spin-projection contribution projected onto the $y$ axis.
From Eq.~\eqref{eq:local-Sy-var}, a local wave packet oriented along $\boldsymbol n$ gives
${\mathrm{Var}_{\boldsymbol n}(S_y)}/{S}=\frac12\left(1-n_y^2\right).$
On the phase-diffusion limit cycle, $n_y=r_{\rm LC}\sin\Phi$, and
$\langle n_y^2\rangle_\Phi=r_{\rm LC}^2\langle\sin^2\Phi\rangle_\Phi={r_{\rm LC}^2}/{2}.$

The two contributions to $\mathrm{Var}_{\rm ss}(S_y)/S$ are therefore
\begin{equation}
	S\mathcal V_{\rm SPN}^{\rm P2}
	=
	\frac12\left(1-\frac{r_{\rm LC}^2}{2}\right)
	=
	\frac{5}{16},
	\qquad
	S\mathcal V_{\rm PDN}^{\rm P2}
	=
	\frac{r_{\rm LC}^2}{2}S
	=
	\frac{3}{8}S.
\end{equation}
Thus P1 and P3 are limited by spin-projection noise in the steady state, whereas P2 and P4 contain an additional phase-diffusion term proportional to $S^2$ in $\mathrm{Var}_{\rm ss}(S_y)$.
This decomposition gives the SPN and PDN contributions shown in Fig.~4(d) of the main text and explains why squeezing does not change the long-time fluctuation of the limit-cycle protocols.

\end{document}